\documentclass[article]{revtex4-1}

\usepackage{amsmath}
\usepackage{amssymb}
\usepackage{graphics}
\usepackage{graphicx}
\usepackage{braket}
\usepackage{hyperref}
\usepackage{color}
\usepackage{xcolor}

\begin{document}


\title{Simulation of Zitterbewegung by modelling the Dirac equation in Metamaterials}


\author{Sven Ahrens}
\email[]{ahrens@csrc.ac.cn}
\affiliation{Beijing Computational Science Research Center, Beijing 100094, China}
\author{Jun Jiang}
\affiliation{Tongji University, Shanghai 200092, China}
\author{Yong Sun}
\affiliation{Tongji University, Shanghai 200092, China}
\author{Shi-Yao Zhu}
\affiliation{Beijing Computational Science Research Center, Beijing 100094, China}


\date{\today}

\begin{abstract}
Strong field processes, which occur in intense laser fields are a test area for relativistic quantum field theory, but difficult to study experimentally and theoretically. Thus, modelling relativistic quantum dynamics and therewith the Dirac equation can help to understand quantum field theory. We develop a dynamic description of an effective Dirac theory in metamaterials, in which the wavefunction is modeled by the corresponding electric and magnetic field in the metamaterial. This electro-magnetic field can be probed in the experimental setup, which means that the wavefunction of the effective theory is directly accessible by measurement. Our model is based on a plane wave expansion, which ravels the identification of Dirac spinors with single-frequency excitations of the electro-magnetic field in the metamaterial. We proof the validity of our relativistic quantum dynamics simulation by demonstrating the emergence of Zitterbewegung and verifying it with an analytic solution.
%
%
%
%
%
%
%
\end{abstract}



\maketitle

\section{Introduction}

Fundamental properties in quantum electro-dynamics can be studied in strong field physics, in which for example intense laser beams can create electron positron pairs in the unstable vacuum\cite{schwinger_1951_pair_creation}. The study of these processes analytically\cite{huayu_carsten_2010_pair_creation} or numerically \cite{hebenstreit_2008_pair_production,matthias_carsten_2009_pair_creation,steinacher_ahrens_grobe_2014_charge_density} demands extended efforts. Therefore modelling systems are discussed in the literature, which allow to re-engineer physics in extreme conditions\cite{szpak_2011_qed_simulator,szpak_2012_qed_simulator}. The capability of simulating intrinsic relativistic quantum processes like pair-creation or Zitterbewegung can therefore improve and test the understanding of quantum field theory.

The latter mentioned Zitterbewegung, first considered by Schr\"odinger \cite{Schroedinger_1930_Zitterbewegung}, is a small quivering motion of particles in relativistic quantum mechanics. The theoretical existence of this quivering motion has been evidenced by numerical simulations of the Dirac equation and of quantum-field-theory\cite{Grobe_1999_Numerical_Dirac_equation_Zitterbewegung,Grobe_2004_Lack_of_Zitterbewegung,Grobe_2008_Zitterbewegung_distinguishable_fermions}. However, the motion itself is very small and very fast. It can be estimated to be at the time scale of electron-positron pair creation ($10^{-20}$ seconds) and has an amplitude on the order of the Compton wavelength of an electron ($10^{-12}$ meters). Therefore, the experimental observation of this counter-intuitive property of nature is challenging.

Nevertheless, the Zitterbewegung can be modeled in various other effective systems and it's appearance has been subject of diverse investigations. It has been discussed for superconductors\cite{Lurie_1970_Zitterbewegung_Superconducting_id_25} in 1970, for one-dimensional chains\cite{Cannata_1990_tight_binding_Zitterbewegung_id_23} in 1990 and many considerations set in on semiconductors\cite{Ferrari_Russo_1990_nonrelativistic_zitterbewegung_in_two-band_systems_id_8,Cannata_Ferrari_1991_Zitterbewegung_two-band_systems_id_9,Zawadzki_2005_zitterbewegung_in_semiductors_id_10,Schliemann_Westervelt_2005_Zitterbewegung_semiconductor_quantum_wells_id_3,Shung-Qing_Shen_2005_zitterbewegung_spin_transverse_force_id_12,Schliemann_Westervelt_2006_Zitterbewegung_semiconductor_quantum_wells_id_11,Rusin_Zawadazki_2007_Zitterbewegung_semiconductors_id_24}. In 2005 renewed discussions on Zitterbewegung in Spintronics\cite{Jiang_2005_zitterbewegung_luttinger_hamiltonian_id_13,Brusheim_2006_electron_waveguide_id_14}, Graphene\cite{Katsnelson_2006_Zitterbewegung_Graphene_id_26,Peres_Novoselov_2009_zitterbewegung_the_electronic_properties_of_graphene_id_18} and carbon nanotubes\cite{Zawadzki_2006_zitterbewegung_carbon_nanotubes_id_15,Tomasz_Wlodek_Zitterbewegung_charge_carriers_graphene_id_5} raised up, (see also\cite{Cserti_2006_zitterbewegung_spintronic_graphene_superconductin_systems_id_1}). Zitterbewegung has also been investigated for trapped ions\cite{Bermudez_2007_zitterbewegung_Jaynes-Cummings_model_id_16,Lamata_2007_zitterbewegung_single_trapped_ion_id_6,Xiong_2008_zitterbewegung_by_quantum_field-theory_considerations_id_17,Wunderlich_2009_cold_trapped_ion_zitterbewegung_id_28}, ultracold atoms\cite{Vaishnav_2008_Zitterbewegung_Ultracold_Atoms_id_4} and Bose-Einstein condensates\cite{Engels_2013_zitterbewegung_bose-einstein_condensate_id_32} with an experimental demonstration of the effect in 2010\cite{Blatt_2010_Zitterbewegung_First_Experiment_id_31}. More considerations on Zitterbewegung aroused for photonic crystals\cite{Zhang_2008_Zitterbewegung_Dirac_Point_id_29}, negative-zero-positive index materials\cite{Zhu_2009_NZPI_Zitterbewegung_id_33} and binary waveguide arrays\cite{Longhi_2010_Zitterbewegung_Binary_Waveguide_Array_id_30}.


However, it is difficult to access the wavefunction itself in the listed experiments, where in particular higher order processes in quantum field theory are at least based on the knowledge of the wavefunction\cite{fradkin_1991_qed_unstable_vacuum}. Recent experimental investigations demonstrated, that the wavefunction of the Dirac equation can be imitated by the electro-magnetic field of a waveguide structure with designable electro-dynamic properties \cite{topological_excitations_tan_2014}. Since this waveguide operates in the microwave regime, the simulated wavefunction can be directly probed in the experiment.  But only quasi-stationary field configurations have been investigated and the question arises, whether the 
time-independent description in terms of frequency-eigensolutions is capable of forming dynamics, which correspond to the Dirac equation.

Here, we extend the quasi-static theory to a dynamic description of an effective Dirac equation  (section \ref{sec:theory_description}) and use it for the demonstration of the Zitterbewegung in theory. In the subsections \ref{sec:maxwell_equations} and \ref{sec:effectice_dirac_equation} we repeat the already established description \cite{topological_excitations_tan_2014} of the Maxwell equations in metamaterials and show how solutions of the Dirac equation can be deduced. Next, we formulate a formal solution of the Maxwell equations by an expansion in plane-wave solutions in subsection \ref{sec:time-evolution_maxwell-equations}. This solution can be identified with the unitary time-evolution of the Dirac equation, which is discussed in subsection \ref{sec:time-evolution_dirac-equation}, in which we also give an explicit mapping between the electro-magnetic field and the Dirac wave-function in frequency- and momentum space. The effective parameters for the mass and the speed of light of the emulated Dirac equation depend on the metamaterial an are derived in subsection \ref{sec:scaled_dirac_equation}. In section \ref{sec:boundary_conditions} we refer back to the Maxwell equations for deriving boundary conditions, with which electro-magnetic input pulses can be injected at the metamaterial interfaces in our simulation.

After the introduction of the theory foundations, we describe the numerical implementation in section \ref{sec:numercial_implementation}, in which we also consider the properties of periodic boundary conditions, which are implied by our simulation method.


In the Results section \ref{sec:results}, we present the metamaterial simulations in three kinds of dynamical scenarios, in which we first consider the easiest possible setup for a Zitterbewegung with a Gaussian wavepacket excitation \ref{sec:gaussian_wavepacket}. This setup is modified into a counterpropagating excitation, such that an experimental realization might be more feasible \ref{sec:counterpropagating_wavepacket} and finally we also account for the influence of the boundaries of the metamaterial \ref{sec:boundary_wavepacket}.

The appendix contains a derivation of a formula of the expectation value of the position expectation operator, with which the Zitterbewegung can be computed semi-analytically\ref{sec:zitterbewegung_of_real_electron}. Another section discusses the absence of imaginary values in a real experiment, while the corresponding Dirac wavefunction consists of complex numbers\ref{sec:imaginary_part}.


\section{Theory description\label{sec:theory_description}}


\subsection{The Maxwell equations\label{sec:maxwell_equations}}

The description of the one-dimensional Maxwell equations
\begin{subequations}
\begin{alignat}{3}
 - \partial_x E_z &= &i &\omega \mu_0 \mu_r(\omega) H_y \label{eq:maxwell_equations_Ez}\\
   \partial_x H_y &= &-i &\omega \epsilon_0 \epsilon_r(\omega) E_z\,.\label{eq:maxwell_equations_Hy}
\end{alignat}\label{eq:maxwell_equations}%
\end{subequations}
in the photonic crystal is adapted from an effective model for the electro-magnetic field in a metamaterial waveguide, which has been developed recently \cite{topological_excitations_tan_2014}. Here, the electric and magnetic field is propagating in the $x$-direction and $\epsilon_0$ and $\mu_0$ are the vacuum permittivity and vacuum permeability, respectively. The frequency dependent relative permittivity and relative permeability is given by
\begin{subequations}
\begin{align}
 \epsilon_r(\omega) &= \frac{1}{p \epsilon_0} \left( C_0 - \frac{1}{\omega^2 L d} \right) \quad \textrm{and}\label{eq:effective_permittivity} \\
 \mu_r(\omega) &= \frac{p}{\mu_0} \left( L_0 - \frac{1}{\omega^2 C d} \right)\,, \label{eq:effective_permeability}
\end{align}\label{eq:permittivity_and_permeability}%
\end{subequations}
where we assume, that damping of the electro-magnetic field is negligible and the metamaterial is homogeneous along the $x$-direction in space. Damping terms may lead to a decay of electro-magnetic excitations, which might be subject to future studies. We assume specific material parameters for our model, which are listed in table \ref{tab:metamaterial_properties}.
\begin{table}[!ht]
\caption{
\bf{Metamaterial parameters}}
\begin{tabular}{ll}
 $d=8\,\textrm{mm}$        & element length\\
 $p=4$                     & geometric factor \\
 $C=2.82\,\textrm{pF}$     & series capacitance of the loading elements \\
 $C_0=58.8\,\textrm{pF/m}$ & per-unit-length capacitance of the trans-\\
                           & mission line segment \\
 $L=19.5\,\textrm{nH}$     & shunt inductance of the loading elements \\
 $L_0=314\,\textrm{nH/m}$  & per-unit-length inductance of the trans-\\
                           & mission line segment
\end{tabular}
\begin{flushleft}
This table contains a list of the parameter properties of the permittivity and permeability in equations \eqref{eq:permittivity_and_permeability}.
\end{flushleft}
\label{tab:metamaterial_properties}
\end{table}
As a result, the electro-magnetic metamaterial properties assume the specific relations
\begin{subequations}
\begin{align}
 \epsilon_r(\omega) &= 1.66 - 188 \left(\frac{\textrm{GHz}}{\omega}\right)^2 \quad \textrm{and}\\
 \mu_r(\omega) &= 1.00 - 141 \left(\frac{\textrm{GHz}}{\omega}\right)^2\,.
\end{align}
\end{subequations}

\subsection{Effective Dirac equation\label{sec:effectice_dirac_equation}}

As discussed in \cite{topological_excitations_tan_2014}, the Maxwell equations can be transformed to a one-dimensional, two-component Dirac equation by introducing
\begin{subequations}
\begin{align}
 \varphi_1 &= \sqrt{\epsilon_0} E_z\,, \\
 \varphi_2 &= \sqrt{\mu_0} H_y\,.
\end{align}\label{eq:wavefunction_em-field_relations}%
\end{subequations}
If one identifies the effective mass
\begin{equation}
 m(\omega) = \frac{\omega}{2 c}\left[ \epsilon_r(\omega) - \mu_r(\omega) \right]\label{eq:Dirac_mass}
\end{equation}
and the effective energy
\begin{equation}
 \mathcal{E}(\omega) = - \frac{\omega}{2 c} \left[ \epsilon_r(\omega) + \mu_r(\omega) \right]\,.\label{eq:Dirac_energy}
\end{equation}
one can rewrite the Maxwell equations, such that they are of the same structural form as the one-dimensional Dirac equation
\begin{equation}
 - i \sigma_x \partial_x \varphi + m(\omega) \sigma_z \varphi = \mathcal{E}(\omega) \varphi \,. \label{eq:metamaterial_dirac_equation}
\end{equation}
Here, $\varphi$ is the two-component wavefunction
\begin{equation}
 \varphi =
 \begin{pmatrix}
  \varphi_1 \\ \varphi_2
 \end{pmatrix}\,,\label{eq:em-wavefunction_transformation}
\end{equation}
$c=1/\sqrt{\epsilon_0 \mu_0}$ is the vacuum speed light and $\sigma_x$ and $\sigma_z$ are the first and third Pauli matrices
\begin{equation}
 \sigma_x =
 \begin{pmatrix}
  0 & 1 \\ 1 & 0
 \end{pmatrix}
 \,,\quad
 \sigma_z =
 \begin{pmatrix}
  1 & 0 \\ 0 & -1
 \end{pmatrix}\,.
\end{equation}

\subsection{Time evolution of Maxwell equations\label{sec:time-evolution_maxwell-equations}}

\begin{figure}
\includegraphics[width=0.48\textwidth]{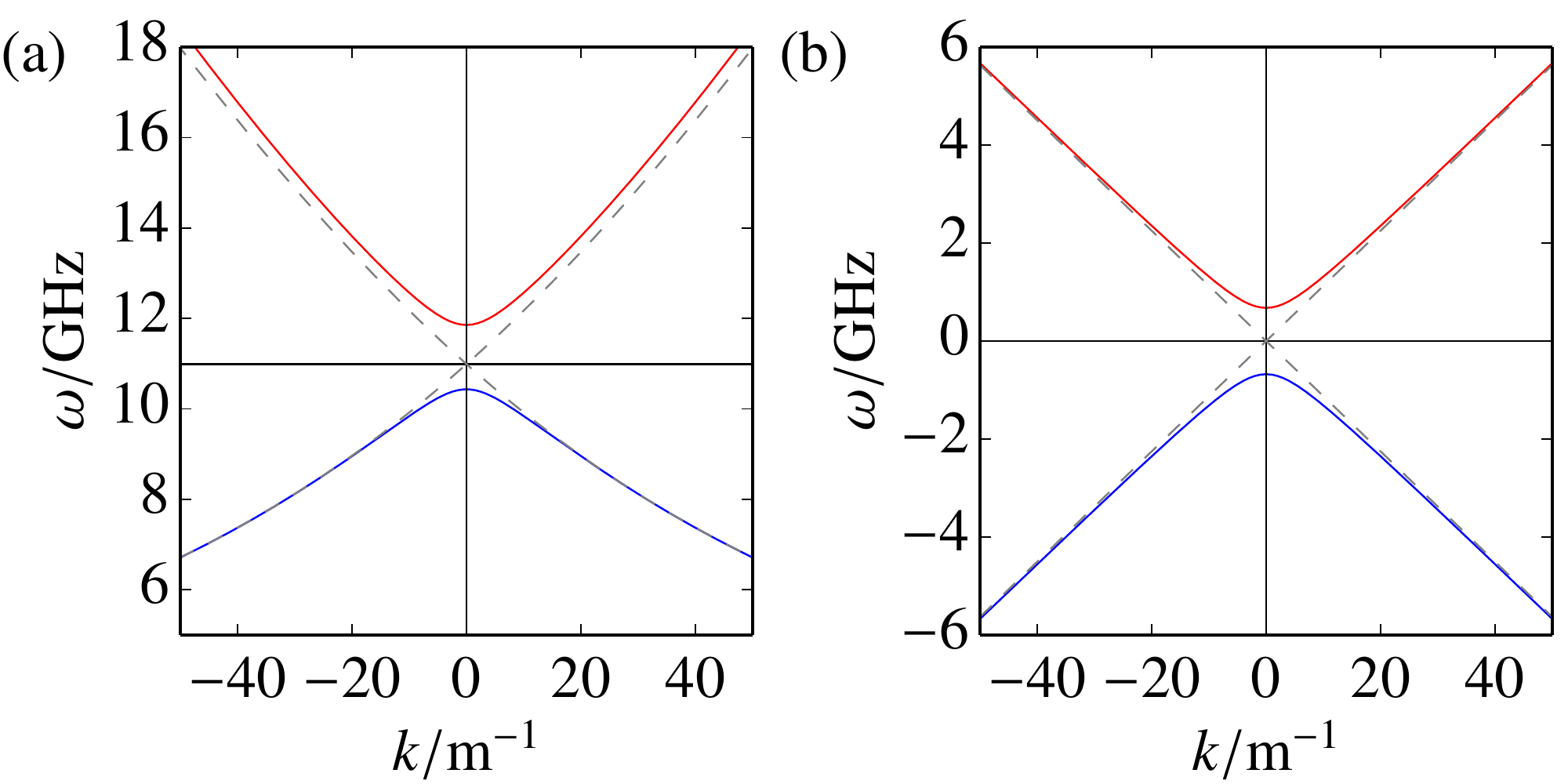}
\caption{(a) Dispersion relation of the metamaterial \eqref{eq:dispersion-relation}, in which the red and blue lines mark the upper band $\omega_+$ and lower band $\omega_-$, respectively. The grey-dashed lines are the functions $\mathcal{E}(\omega)$ and $-\mathcal{E}(\omega)$, which are plotted along the $x$-axis as functions of the frequency on the $y$-axis. (b) The same plot as in (a), but for the relativistic energy-momentum relation \eqref{eq:relativistic_energy-momentum-relation} of the scaled, exact Dirac theory. The gray-dashed lines exhibit the scaled light cone $|\omega|= c_D|k|$.}
\label{fig:dispersion_relation}
\end{figure}

In order to establish a dynamic Dirac theory, we first consider the time-evolution of the Maxwell equations \eqref{eq:maxwell_equations}. Since the Maxwell equations are specified in frequency space, we evolve the electro-magnetic field in the metamaterial by the time-evolution of an expansion of  plane wave momentum eigenfunctions $e^{i k x}$. Applying these plane waves yields the Maxwell equations in frequency- and momentum-space
\begin{subequations}
\begin{align}
 k E_{z,k}(\omega) &= - \omega \mu_0 \mu_r(\omega) H_{y,k}(\omega) \label{eq:maxwell_equations_frequency_space_E_z_k}\,,\\
 k H_{y,k}(\omega) &= - \omega \epsilon_0 \epsilon_r(\omega) E_{z,k}(\omega) \label{eq:maxwell_equations_frequency_space_H_y_k}\,,
\end{align} \label{eq:maxwell_equations_frequency_space_k}%
\end{subequations}
where the $E_{z,k}(\omega)$ and $H_{y,k}(\omega)$ are the electric and magnetic field amplitudes for a certain field frequency $\omega$ and a certain momentum $k$. From these Maxwell equations one can derive the dispersion relation
\begin{equation}
 k^2 = \frac{\omega^2}{c^2} \epsilon_r(\omega) \mu_r(\omega)\,. \label{eq:dispersion-relation}
\end{equation}
The dispersion relation implies, that there exists a positive and negative momentum $k$ for each frequency $\omega$. Furthermore, if one inserts the permittivity \eqref{eq:effective_permittivity} and permeability \eqref{eq:effective_permeability}, one obtains a polynomial of second order in $\omega^2$. This means, that there exist two frequencies (upper band $\omega_+$ and lower band $\omega_-$) and the identical negative values $-\omega_+$ and $-\omega_-$ for each momentum $k$. This is plotted in figure \ref{fig:dispersion_relation} (a) for the two positive frequencies $\omega_+$ and $\omega_-$.
Upper and lower band are divided by the band gap in which the product of $\epsilon_r(\omega) \mu_r(\omega)$ turns negative. Outside of the band gap, the product $\epsilon_r(\omega) \mu_r(\omega)$ is always positive, and if also $k^2$ is positive (ie. $k\in\mathbb{R}$), it follows, that $\omega^2$ is positive too and therefore $\omega \in \mathbb{R}$. The upper and lower boundary of the band gap in frequency space is given by the frequencies, at which $\epsilon_r$ and $\mu_r$ are zero. This happens at the frequency
\begin{equation}
 \omega_1 = 10.4\,\textrm{GHz}\,,
\end{equation}
at which $\epsilon_r$ is zero and at the frequency
\begin{equation}
 \omega_2 = 11.8\,\textrm{GHz}\,,
\end{equation}
at which $\mu_r$ is zero, for the parameters of table \ref{tab:metamaterial_properties}. The comparison of the dispersion relation of the metamaterial in Fig. \ref{fig:dispersion_relation} (a) with the relativistic energy-momentum relation of the scaled, exact Dirac theory in Fig. \ref{fig:dispersion_relation} (b) illustrates a possible dynamical similarity of both descriptions around the momentum $k=0\,\textrm{m}^{-1}$.

Since the Maxwell equations are linear differential equations, the superposition of solutions is a solution again. Therefore, the time-evolution of the electro-magnetic field can be written down by the expansion
\begin{subequations}
\begin{align}
 E_z(x,t) &= \sum_{k \atop{\{\omega_+,\omega_-,\atop{-\omega_+,-\omega_-\}}}} E_{z,k}(\omega) e^{i (k x - \omega t)} \ \Delta k\,, \label{eq:E-field_expansion}\\
 H_y(x,t) &= \sum_{k \atop{\{\omega_+,\omega_-,\atop{-\omega_+,-\omega_-\}}}} H_{y,k}(\omega) e^{i (k x - \omega t)} \ \Delta k\,, \label{eq:H-field_expansion}
\end{align}\label{eq:field_expansion}%
\end{subequations}
in which the sum runs over all discrete momenta $k$ and all frequencies $\{\omega_+(k),\omega_-(k),-\omega_+(k),-\omega_-(k)\}$ for each momentum. The factor $\Delta k$ is a measure for the summation in momentum space and is considered to be analogous to the measure $dk$ of an integral. Since momentum space is spaced equidistantly in our approach, we write down a sum in \eqref{eq:field_expansion} with measure $\Delta k$ and focus on the plane wave solution in this more fundamental theory part. The spacing of $\Delta k$ and it's relation to position space is discussed in the section on numerical implementation \ref{sec:numercial_implementation}.

In the expansion \eqref{eq:field_expansion}, the magnetic field can be expressed by the electric field by using one of the two Maxwell equations \eqref{eq:maxwell_equations_frequency_space_k}, yielding
\begin{subequations}
\begin{align}
 H_{y,k}(\omega) &= - \frac{k}{\omega \mu_0 \mu_r(\omega)} E_{z,k}(\omega)\textrm{, or}\label{eq:E-H-relation_mu} \\
 H_{y,k}(\omega) &= - \frac{\omega \epsilon_0 \epsilon_r(\omega)}{k} E_{z,k}(\omega)\,.\label{eq:E-H-relation_epsilon}
\end{align}\label{eq:E-H-relation}%
\end{subequations}
Therefore, the sum \eqref{eq:E-field_expansion} is already sufficient for a unique specification of the electro-magnetic field in the metamaterial.

\subsection{Time evolution of the effective Dirac equation\label{sec:time-evolution_dirac-equation}}

In order to write down the time-evolution in Dirac theory, the plane wave solutions of the Dirac equation \eqref{eq:metamaterial_dirac_equation} are to be considered. These are usually obtained by applying the plane waves $e^{i k x}$ at the Hamiltonian and solving the eigenvector and eigenvalue problem of the Hamiltonian matrix. The left-hand side of the Dirac equation \eqref{eq:metamaterial_dirac_equation} turns into the matrix
\begin{equation}
\begin{pmatrix}
m(\omega) & k \\
k & - m(\omega)
\end{pmatrix}\,,\label{eq:dirac_matrix}
\end{equation}
when applying the plane wave $e^{i k x}$. The definition of the mass \eqref{eq:Dirac_mass} and energy \eqref{eq:Dirac_energy} together with the dispersion relation \eqref{eq:dispersion-relation} ensure the identity
\begin{equation}
 \mathcal{E}(\omega)^2 = k^2 + m(\omega)^2\,,\label{eq:energy_momentum_relation}
\end{equation}
which can be seen as relativistic energy-momentum relation of an electro-magnetic excitation in the metamaterial with wave number $k$ and corresponding frequency $\omega$. Making use of \eqref{eq:energy_momentum_relation} one obtains the eigenvector
\begin{subequations}
\begin{equation}
 u_{k}^+(\omega) =
 \frac{1}{\sqrt{2 |\mathcal{E}(\omega)|[|\mathcal{E}(\omega)| + m(\omega)]}}
 \begin{pmatrix}
  |\mathcal{E}(\omega)| + m(\omega) \\ k
 \end{pmatrix}\label{eq:positive_spinor}
\end{equation}
with eigenvalue $|\mathcal{E}(\omega)|$ and the eigenvector
\begin{equation}
 u_{k}^-(\omega) =
 \frac{1}{\sqrt{2 |\mathcal{E}(\omega)|[|\mathcal{E}(\omega)| + m(\omega)]}}
 \begin{pmatrix}
  - k \\ |\mathcal{E}(\omega)| + m(\omega)
 \end{pmatrix}\label{eq:negative_spinor}
\end{equation}\label{eq:spinors}%
\end{subequations}
with eigenvalue $-|\mathcal{E}(\omega)|$, for the matrix \eqref{eq:dirac_matrix}. The solutions \eqref{eq:spinors} are commonly known as bi-spinors of the Dirac equation in one dimension \cite{parthasarathy_2010_rel_quantum_mech}. We have normalized the bi-spinors in our notion, such that they form an an orthonormal base. Note, that the spinors \eqref{eq:spinors} have the identical analytic expressions
\begin{subequations}
\begin{equation}
 u_{k}^+(\omega) =
 \frac{\textrm{sign}(k)}{\sqrt{2 |\mathcal{E}(\omega)|[|\mathcal{E}(\omega)| - m(\omega)]}}
 \begin{pmatrix}
  k \\ |\mathcal{E}(\omega)| - m(\omega)
 \end{pmatrix}\label{eq:positive_spinor_alternative}
\end{equation}
and
\begin{equation}
 u_{k}^-(\omega) =
 \frac{\textrm{sign}(k)}{\sqrt{2 |\mathcal{E}(\omega)|[|\mathcal{E}(\omega)| - m(\omega)]}}
 \begin{pmatrix}
  - |\mathcal{E}(\omega)| + m(\omega) \\ k
 \end{pmatrix}\label{eq:negative_spinor_alternative}
\end{equation}\label{eq:spinor_alternative}%
\end{subequations}
respectively, with the signum of $k$
\begin{equation}
 \textrm{sign}(k) :=
 \begin{cases}
  \phantom{-}1 & \textrm{, if } 0 \le k \\
  -1 & \textrm{\, else.}
 \end{cases}
\end{equation}
Even though the eigensolutions of the matrix \eqref{eq:dirac_matrix} are determined by (\ref{eq:spinors},\ref{eq:spinor_alternative}), one has to keep in mind, that this matrix is an integrated part of equation \eqref{eq:metamaterial_dirac_equation}. As a consequence the plane wave eigensolutions are formed by the electro-dynamics in the metamaterial. Since $\mathcal{E}(\omega)$ is negative in the upper band and positive in the lower band, only the negative eigensolution $u_k^- e^{i k x}$ appears in the upper band and only the positive eigensolution $u_k^+ e^{i k x}$ appears in the lower band. In fact, the spinors \eqref{eq:spinors} can be expressed in terms of the electro-magnetic field. Keeping in mind, that $\mathcal{E}(\omega)$ is positive at the lower band, the spinor $u_k^+$ can be written as
\begin{subequations}
\begin{align}
 u_{k}^+(\omega_-) &=
 \frac{1}{\sqrt{2 \mathcal{E}(\omega_-)[\mathcal{E}(\omega_-) + m(\omega_-)]}}
 \begin{pmatrix}
  \mathcal{E}(\omega_-) + m(\omega_-) \\ k
 \end{pmatrix}\nonumber\\
 &=\frac{1}{\sqrt{[\epsilon_r(\omega_-) + \mu_r(\omega_-)] \mu_r(\omega_-)}}
 \begin{pmatrix}
  - \mu_r(\omega_-) \\ \frac{c k}{\omega_-}
 \end{pmatrix} \,.\label{eq:positive_spinor_em}
\end{align}
Similarly, $\mathcal{E}(\omega)$ is negative at the upper band and by using the solution \eqref{eq:negative_spinor_alternative} for $u_k^-$, one can write
\begin{align}
 u_{k}^-(\omega_+) &=
 \frac{\textrm{sign}(k)}{\sqrt{2 \mathcal{E}(\omega_+)[\mathcal{E}(\omega_+) + m(\omega_+)]}}
 \begin{pmatrix}
  \mathcal{E}(\omega_+) + m(\omega_+) \\ k
 \end{pmatrix}\nonumber\\
 &=\frac{\textrm{sign}(k)}{\sqrt{[\epsilon_r(\omega_+) + \mu_r(\omega_+)] \mu_r(\omega_+)}}
 \begin{pmatrix}
  - \mu_r(\omega_+) \\ \frac{c k}{\omega_+}
 \end{pmatrix} \,,\label{eq:negative_spinor_em}
\end{align}\label{eq:spinor_em}%
\end{subequations}
which is identical to \eqref{eq:positive_spinor_em}, except the sign of $k$. The $\textrm{sign}(k)$ factor in \eqref{eq:negative_spinor_em} is necessary, because we want to keep the convention of the spinor solutions \eqref{eq:spinors}, as close to common relativistic quantum dynamics as possible (see also \eqref{eq:spinors_vacuum}). Note, that the fraction of upper and lower component of the spinors \eqref{eq:spinor_em} corresponds to the factor \eqref{eq:E-H-relation_mu} between electric and magnetic field, if one accounts for the transformation law \eqref{eq:wavefunction_em-field_relations} of the electro-magnetic field of the Dirac wave function. Therefore, the free eigensolutions of the Dirac equation are linearly dependent to the plane wave solutions of the Maxwell equations and it just remains to determine the proportionality constant, to make a complete, unique link between the Maxwell equations of the electro-magnetic field in the metamaterial and the corresponding dynamic Dirac equation. To do so, we write down the time evolution of the wave function by expanding it with respect to the obtained eigenfunctions $u_k^+ e^{i k x}$ and $u_k^- e^{i k x}$ of the Dirac equation \eqref{eq:metamaterial_dirac_equation}.
\begin{equation}
 \varphi(x,t) = \frac{1}{\sqrt{N}}\sum_{k} \Bigg[\phi_k^+(\omega_-) \, u_k^+(\omega_-) \, e^{i (k x - \omega_- t)}
 + \phi_k^-(\omega_+) \, u_k^-(\omega_+) \, e^{i (k x - \omega_+ t)}\Bigg] \Delta k\,,\label{eq:wavefunction_expansion}
\end{equation}
Here, $\phi_k^+(\omega)$ and $\phi_k^-(\omega)$ are expansion coefficients,
\begin{equation}
 N=\int_{-\infty}^\infty |\varphi(x,t)|^2 dx \label{eq:position_space_normalization}
\end{equation}
is a normalization constant and $\Delta k$ is the measure for the momentum space summation, which is already mentioned in subsection \ref{sec:time-evolution_maxwell-equations} and further discussed in section \ref{sec:numercial_implementation}. Since \eqref{eq:wavefunction_expansion} is a unitary transformation of the initial state of the wave function, the normalization constant will keep constant in time and $\varphi(x,t)$ will always stay normalized at one. Note, that according to the considerations above, the positive eigensolutions which are proportional to $u_k^+$ will evolve with the lower band frequency $\omega_-$ and the negative eigensolutions which are proportional to $u_k^-$ will evolve with the upper band frequency $\omega_+$.

If one compares the expansion of the electric field \eqref{eq:E-field_expansion} with the first component of the wave function \eqref{eq:wavefunction_expansion} and requires equality for the prefactors of the exponential $e^{i k x - \omega_+}$ one obtains
\begin{subequations}
\begin{equation}
 E_{z,k}(\omega_+) = - \phi_k^-(\omega_+) \frac{\textrm{sign}(k) c \sqrt{\mu_0 |\mu_r(\omega_+)|}}{\sqrt{|\epsilon_r(\omega_+) + \mu_r(\omega_+)|}}\,. \label{eq:expansion_coefficient_equality_upper_band}
\end{equation}
Similarly, one obtains
\begin{equation}
 E_{z,k}(\omega_-) = \phi_k^+(\omega_-) \frac{c \sqrt{\mu_0 |\mu_r(\omega_-)|}}{\sqrt{|\epsilon_r(\omega_-) + \mu_r(\omega_-)|}} \label{eq:expansion_coefficient_equality_lower_band}
\end{equation}\label{eq:expansion_coefficient_equality}%
\end{subequations}
if one requires equality for the prefactors of $e^{i k x - \omega_-}$. One may also equalize the expansion of the magnetic field \eqref{eq:H-field_expansion} with the second component of the wave function \eqref{eq:wavefunction_expansion} for the exponentials $e^{i k x - \omega_+}$ and $e^{i k x - \omega_-}$. The result will be equivalent to equations \eqref{eq:expansion_coefficient_equality}, because the spinors \eqref{eq:spinor_em} are linearly dependent on the electro-magnetic excitations in the expansions \eqref{eq:field_expansion}.

\subsection{The metamaterial's scaled Dirac equation\label{sec:scaled_dirac_equation}}

According to the considerations in the sections \ref{sec:time-evolution_maxwell-equations} and \ref{sec:time-evolution_dirac-equation} the metamaterial's emulated quantum dynamics is expected to evolve as in Dirac theory, but natural constants like the electron mass and the effective speed of light depend on the metamaterial properties.

For deducing these properties, we first consider the time-evolution equation
\begin{equation}
 i \hbar \partial_t \varphi(x,t) = H \varphi(x,t)\,,\label{eq:quantum_mechanical_time_evolution}
\end{equation}
with the Dirac Hamiltonian
\begin{equation}
 H = c p_x \sigma_x + m_0 c^2 \sigma_z\,,\label{eq:vacuum_dirac_hamiltonian}
\end{equation}
in relativistic quantum mechanics. Here, $m_0$ is the electron rest mass and 
\begin{equation}
 p_x=- i \hbar \partial_x \label{eq:momentum_operator}
\end{equation}
is the momentum operator. After a Fourier transform of this equation into frequency space, the time-derivative $i \partial_t$ is substituted by $\omega$, yielding
\begin{equation}
 H \varphi(x,\omega) = \hbar \omega \varphi(x,\omega)\,.\label{eq:desired-dirac_form}
\end{equation}
If equation \eqref{eq:metamaterial_dirac_equation} could be cast in a similar shape with a linear $\omega$ at the right-hand side, it's generalization to the time-domain would be straight forward and the mass and speed of light parameters of the emulated Dirac equation could be read of from the corresponding Hamiltonian on the left-hand side. 
%
%
%
%
This can be achieved by a Taylor expansion of the effective Energy $\mathcal{E}(\omega)$ and effective mass $m(\omega)$ of the metamaterial Dirac equation \eqref{eq:metamaterial_dirac_equation}, at the frequency $\omega_0$, at which $\mathcal{E}(\omega)$ is zero. Thus, $\omega_0$ is determined by setting the definition \eqref{eq:Dirac_energy} of $\mathcal{E}(\omega)$ to zero and solving for $\omega$, resulting in
\begin{equation}
 \omega_0 = \sqrt{\frac{1}{CLd}\frac{\mu_0 C + p^2 \epsilon_0 L}{\mu_0 C_0 + p^2 \epsilon_0 L_0}}\,.
\end{equation}
For the parameters chosen in table \ref{tab:metamaterial_properties} $\omega_0$ has the value $11.0\,\textrm{GHz}$. The first relevant order of the Taylor expansion is sufficient for our consideration. For the mass $m(\omega)$, the first relevant order is the zeroth order, whereas for the effective Energy $\mathcal{E}(\omega)$, the first relevant order is the first order. The reason is, that $\omega_0$ is chosen such that the zeroth order of $\mathcal{E}(\omega)$ vanishes. Accordingly, one obtains
\begin{equation}
 - i \sigma_x \partial_x \varphi + m(\omega_0) \sigma_z \varphi = \left. \frac{\partial \mathcal{E}}{\partial \omega}\right|_{\omega_0} (\omega - \omega_0) \,\varphi \label{eq:taylored_dirac_equation}
\end{equation}
for equation \eqref{eq:metamaterial_dirac_equation}. If one defines the shifted frequency
\begin{equation}
 \Delta \omega := \omega - \omega_0\,,
\end{equation}
the effective speed of light
\begin{equation}
 c_D := - \left(\left.\frac{\mathcal{E}(\omega)}{\partial \omega}\right|_{\omega_0}\right)^{-1},
\end{equation}
the new effective mass
\begin{equation}
 m' = m(\omega_0)\frac{\hbar}{c_D}
\end{equation}
and multiplies equation \eqref{eq:taylored_dirac_equation} with $\hbar\, c_D$, then equation \eqref{eq:taylored_dirac_equation} can be transformed into
\begin{equation}
 p_x c_D \sigma_x \varphi + m' c_D^2 \sigma_z \varphi = - \hbar \, \Delta \omega \,\varphi\,.\label{eq:scaled_dirac_equation}
\end{equation}
This is consistent with the Hamiltonian in equation \eqref{eq:desired-dirac_form} and a negative time evolution, according to equation \eqref{eq:quantum_mechanical_time_evolution}. Comparing the corresponding Hamiltonian
\begin{equation}
 H = p_x c_D \sigma_x + m' c_D^2 \sigma_z
\end{equation}
with the Hamiltonian of Dirac theory \eqref{eq:vacuum_dirac_hamiltonian}, one deduces the substitution
\begin{subequations}
\begin{align}
 c &\rightarrow c_D\,, \\
 m_0 &\rightarrow m'\,,
\end{align}\label{eq:scaling_replacements}%
\end{subequations}
which scales the Dirac theory of electrons such that it is similar to the emulated Dirac dynamics of the metamaterial.

We point out, that equation \eqref{eq:scaled_dirac_equation} implies a negative time-evolution, because it's right-hand side contains a minus sign in front of the $\hbar \Delta \omega \varphi$ expression. In a Dirac equation with positive time-evolution, this minus sign is a plus sign (see equation \eqref{eq:desired-dirac_form}). The minus sign originates from our chosen convention (\ref{eq:wavefunction_em-field_relations},\ref{eq:Dirac_mass},\ref{eq:Dirac_energy}). A different convention might require an inversion of the $x$-axis. Therefore, if we want to compare the metamaterial dynamics with results from exact Dirac theory, we have to revert time. In particular we have to add a minus sign in the time argument of the semi-analytic expression \eqref{eq:zitterbewegung_expectation_complex} and \eqref{eq:zitterbewegung_expectation} of the Zitterbewegung, which is derived in appendix \ref{sec:zitterbewegung_of_real_electron}.

\section{Boundary conditions\label{sec:boundary_conditions}}

In this section, we consider electro-dynamic boundary conditions, for accounting for the finite space extension of the metamaterial. We assume, that the 'left' end of the metamaterial is at location $x_a$ and the 'right' end of the metamaterial is at location $x_b$, where the notion 'left' and 'right' implies that $x_a < x_b$ on the $x$-coordinate. As a consequence, the positions $x_a$ and $x_b$ are the limiters of three different regions, which are denoted by the indices $(1)$, $(2)$ and $(3)$, in which the physical space is divided into
\begin{align}
 x < x_a &:\quad \textrm{region (1), the left input wave guide,} \nonumber \\
 x_a < x < x_b &:\quad \textrm{region (2), the metamaterial,} \\
 x_b < x \phantom{< x_a,}&:\quad \textrm{region (3), the right input wave guide.} \nonumber
\end{align}
The boundary conditions are obtained by integration of equation \eqref{eq:maxwell_equations} over the infinitesimal interval $[x_a - \epsilon,x_a + \epsilon]$ and $[x_b - \epsilon,x_b + \epsilon]$.
We make the reasonable assumption, that the right-hand side of equation \eqref{eq:maxwell_equations} is bound and has a finite value. In this case the integrals over the intervals with infinitely small $\epsilon$ imply, that the electric and magnetic fields must be smooth at the boundaries $x_a$ and $x_b$, which means that
\begin{subequations}
\begin{align}%
 E_z^{(1)}(x_a,\omega) &= E_z^{(2)}(x_a,\omega)\,,\\
 E_z^{(2)}(x_b,\omega) &= E_z^{(3)}(x_b,\omega)\,,\\
 H_y^{(1)}(x_a,\omega) &= H_y^{(2)}(x_a,\omega)\,,\\
 H_y^{(2)}(x_b,\omega) &= H_y^{(3)}(x_b,\omega)%
\end{align}\label{eq:E-and-H_continuity_omega}%
\end{subequations}
holds.

We assume that the relative permittivity and permeability are just $1$ for the propagation along the input wave guides of the regions (1) and (3), like it is for electro-magnetic waves in vacuum. Then, the dispersion relation \eqref{eq:dispersion-relation} simplifies to
\begin{equation}
 |\omega| = c |k|.
\end{equation}

For left propagating waves, in which $E_{z,k}(\omega) = H_{y,k}(\omega) = 0$ and $E_{z,-k}(-\omega) = H_{y,-k}(-\omega) = 0$, the field expansion \eqref{eq:field_expansion} can be simplified to
\begin{subequations}
\begin{align}
 E_{z,\textrm{left}}(x,t) &= \sum_{k , \{\omega,-\omega\}} E_{z,k}(\omega) e^{i k (x - c t)}\,,\\
 H_{y,\textrm{left}}(x,t) &= \sum_{k ,\{\omega,-\omega\}} H_{y,k}(\omega) e^{i k (x - c t)}\,,
\end{align}\label{eq:left-moving_field_expansion}%
\end{subequations}
for the regions (1) and (3). Similarly, for right propagating waves, in which $E_{z,-k}(\omega) = H_{y,-k}(\omega) = 0$ and $E_{z,k}(-\omega) = H_{y,k}(-\omega) = 0$, the field expansion \eqref{eq:field_expansion} can be simplified to
\begin{subequations}
\begin{align}
 E_{z,\textrm{right}}(x,t) &= \sum_{k ,\{\omega,-\omega\}} E_{z,k}(\omega) e^{i k (x + c t)}\,,\\
 H_{y,\textrm{right}}(x,t) &= \sum_{k ,\{\omega,-\omega\}} H_{y,k}(\omega) e^{i k (x + c t)}\,,
\end{align}\label{eq:right-moving_field_expansion}%
\end{subequations}
for the regions (1) and (3). The simplified expansions imply a dispersionless translation of the signal
\begin{subequations}
\begin{align}
 E_{z,\textrm{left}}(x,t) &= E_{z,\textrm{left}}(x + c \Delta t,t + \Delta t) \,,\\
 E_{z,\textrm{right}}(x,t) &= E_{z,\textrm{right}}(x - c \Delta t,t + \Delta t) \,,\\
 H_{z,\textrm{left}}(x,t) &= H_{z,\textrm{left}}(x + c \Delta t,t + \Delta t) \,,\\
 H_{z,\textrm{right}}(x,t) &= H_{z,\textrm{right}}(x - c \Delta t,t + \Delta t)
\end{align}\label{eq:signal_displacement}%
\end{subequations}
in the regions (1) and (3). In other words: Once the left and right propagating input and output signals are known in the regions (1) and (3) at any position $x$, they can be trivially deduced from the above equations. It is most convenient to know the input signal at the metamaterial boundaries $x_a$ and $x_b$. Therefore, we introduce the new coordinates
\begin{subequations}
\begin{align}
 x' &:= x - x_a \quad \textrm{for region (1) and}\\
 x^{\prime\prime} &:= x - x_b \quad\, \textrm{for region (3)}\,.
\end{align}%
\end{subequations}
The conditions \eqref{eq:E-and-H_continuity_omega} change into
\begin{subequations}
\begin{align}%
 E_z^{(1)}(0,\omega) &= E_z^{(2)}(x_a,\omega)\,,\\
 E_z^{(2)}(x_b,\omega) &= E_z^{(3)}(0,\omega)\,,\\
 H_y^{(1)}(0,\omega) &= H_y^{(2)}(x_a,\omega)\,,\\
 H_y^{(2)}(x_b,\omega) &= H_y^{(3)}(0,\omega)%
\end{align}%
\end{subequations}
in terms of these new coordinates. If one inserts the field expansion \eqref{eq:field_expansion} in these continuity conditions and keeps in mind, that the exponentials $e^{-i \omega t}$ are linearly independent on the time interval $]-\infty,\infty[$ for each frequency $\omega$, one obtains the boundary conditions in frequency space
\begin{subequations}
\begin{equation}
E_{z,k}^{(1)}(\omega) + E_{z,-k}^{(1)}(\omega) = E_{z,k}^{(2)}(\omega) e^{i k x_a} + E_{z,-k}^{(2)}(\omega) e^{- i k x_a}\label{eq:explicit_boundary_conditions_left_E}
\end{equation}
\begin{equation}
 E_{z,k}^{(2)}(\omega) e^{i k x_b} + E_{z,-k}^{(2)}(\omega) e^{- i k x_b} = E_{z,k}^{(3)}(\omega) + E_{z,-k}^{(3)}(\omega) \label{eq:explicit_boundary_conditions_right_E}\\
\end{equation}
\begin{equation}
 H_{y,k}^{(1)}(\omega) + H_{y,-k}^{(1)}(\omega) = H_{y,k}^{(2)}(\omega) e^{i k x_a} + H_{y,-k}^{(2)}(\omega) e^{- i k x_a}\label{eq:explicit_boundary_conditions_H_left}
\end{equation}
\begin{equation}
 H_{y,k}^{(2)}(\omega) e^{i k x_b} + H_{y,-k}^{(2)}(\omega) e^{- i k x_b} =  H_{y,k}^{(3)}(\omega) + H_{y,-k}^{(3)}(\omega) \,.\label{eq:explicit_boundary_conditions_H_right}
\end{equation}\label{eq:explicit_boundary_conditions}%
 \end{subequations}
The relations \eqref{eq:E-H-relation} allow for the substitution of the magnetic field expansion coefficients in \eqref{eq:explicit_boundary_conditions_H_left} and \eqref{eq:explicit_boundary_conditions_H_right} with the electric field expansion coefficients, resulting in 
\begin{subequations}
\begin{equation}
 -E_{z,k}^{(1)}(\omega) \frac{1}{c \mu_0} + E_{z,-k}^{(1)}(\omega) \frac{1}{c \mu_0} = E_{z,k}^{(2)}(\omega) F^{(2)}_k(\omega) e^{i k x_a} + E_{z,-k}^{(2)}(\omega) F^{(2)}_{-k}(\omega) e^{- i k x_a}\,,
\end{equation}
\begin{equation}
 E_{z,k}^{(2)}(\omega) F^{(2)}_k(\omega) e^{i k x_b} + E_{z,-k}^{(2)}(\omega) F^{(2)}_{-k}(\omega) e^{- i k x_b} = -E_{z,k}^{(3)}(\omega) \frac{1}{c \mu_0} + E_{z,-k}^{(3)}(\omega) \frac{1}{c \mu_0}\,,
\end{equation}\label{eq:explicit_boundary_conditions_II}
\end{subequations}
where
\begin{equation}
 F^{(2)}_k(\omega) = - \frac{k}{\omega \mu_0 \mu_r^{(2)}(\omega)} = - \frac{\omega \epsilon_0 \epsilon_r^{(2)}(\omega)}{k}
\end{equation}
is an abbreviation for convenience.

\section{Numerical implementation\label{sec:numercial_implementation}}

The numerical implementation makes use of an equidistant grid of states in momentum space, which by Fourier transform also implies an equidistant grid in position space. Assume, there are $n$ grid points, each of them spaced by the extension $\Delta x=8\,\textrm{mm}$ of the metamaterial's unit element size (see table \ref{tab:metamaterial_properties}). This implies a length $l$ of the metamaterial, which is $l=n \,\Delta x$. Since the plane waves $e^{i k x}$ should be $2\pi$ periodic after this extension $l$, the spacing in momentum space must be $\Delta k = 2 \pi/l$. Thus, in momentum space, the momenta reach from $-\Delta k (n-1)/2$ till $\Delta k (n-1)/2$, where $n$ must be an odd number. The advantage of an equidistant momentum spacing is, that the plane wave spinors $u_k^{\pm} e^{i k x}$ in the expansion \eqref{eq:wavefunction_expansion} are complete and orthonormal basis functions on the spacial interval $[-l/2 , l/2]$. Furthermore, the simulation has periodic boundary conditions at the interval limits, because all exponentials are periodically continuing. Therefore, the simulations in the subsections \ref{sec:gaussian_wavepacket} and \ref{sec:counterpropagating_wavepacket} are valid, as long as the wavefunction's probability density stays within the boundaries of the simulation area.

\begin{figure}
\begin{center}
\includegraphics[width=0.45\textwidth]{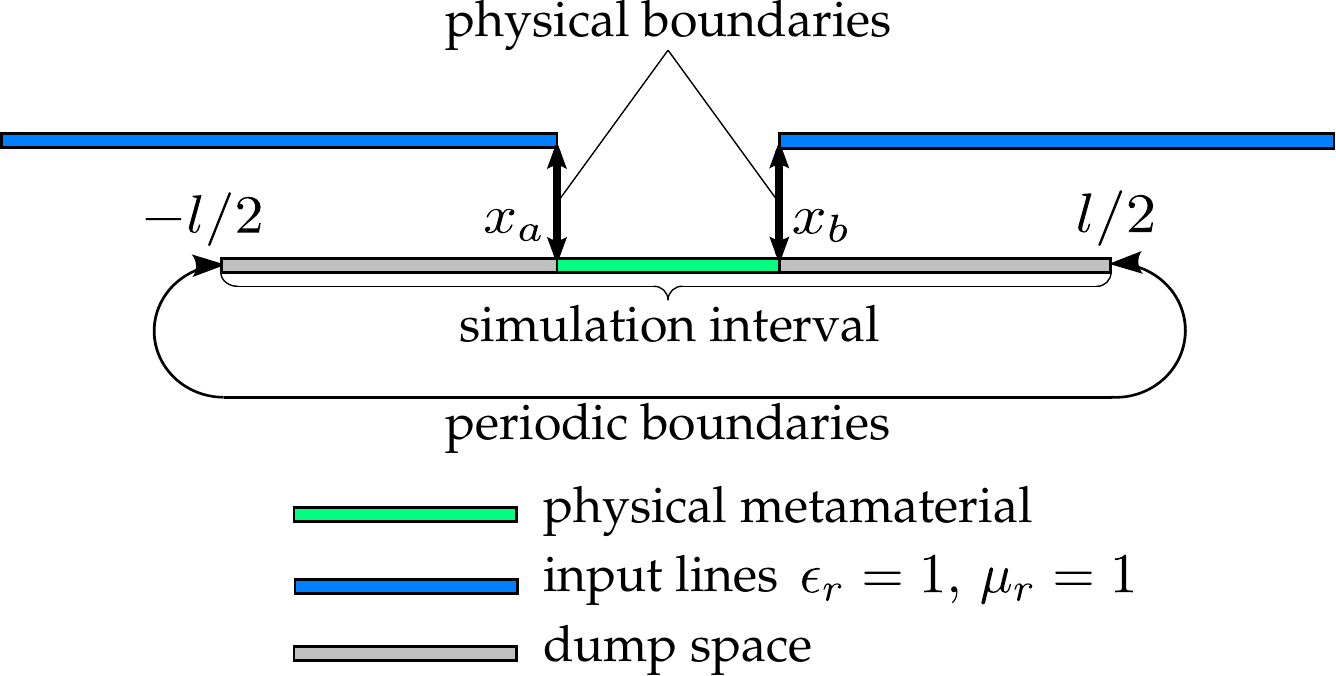}
\end{center}
\caption{
Implementation of boundary conditions within the periodic simulation area in subsection \ref{sec:boundary_wavepacket}. The unitary time-evolution according to \eqref{eq:wavefunction_expansion} implicitly involves periodic boundary conditions. Therefore, the physical boundary conditions according to section \ref{sec:boundary_conditions} are invoked at the space in-between the periodic boundaries of the whole simulation area. Since the simulated metamaterial slab is embedded within some dump space, both boundary conditions are not interfering with each other. The left input line is extended infinitely to the left and the right input line is extended infinitely to the right, respectively. They are described by the equations (\ref{eq:left-moving_field_expansion},\ref{eq:right-moving_field_expansion},\ref{eq:signal_displacement}) of region (1) and (3). The green region of the physical metamaterial corresponds to region (2).
}
\label{fig:periodic_boundary_conditions}
\end{figure}

In subsection \ref{sec:boundary_wavepacket} the initial condition of the simulation is not specified at an initial time, but imposed by additional boundary conditions instead. These boundary conditions, which are discussed in section \ref{sec:boundary_conditions}, are modeling the injection of input pulses at the physical boundaries of the metamaterials. However, the positions $x_a$ and $x_b$, at which the injection boundary conditions are imposed are within the simulation area (see illustrative figure \ref{fig:periodic_boundary_conditions}), such that they don't interfere with each other.

The time-evolution is computed according to equation \eqref{eq:wavefunction_expansion}, which is implemented numerically by first multiplying each expansion coefficient $\phi_k$ with it's time-evolving phase factor $e^{- i \omega t}$ at the certain time $t$ and then summing over $k$. Since the sum over $k$ runs over the exponentials $e^{i k x}$ and the discrete set of momenta $k$ is equidistantly spaced, the sum is executed by performing a Fourier transformation of the $\phi_k \, e^{- i \omega(k) t}$ array.

\section{Results from simulation\label{sec:results}}

In this results section, we apply the theory considerations from above, for demonstrating a Zitterbewegung of the emulated Dirac dynamics in the metamaterial. In a series of three different simulations, we first consider a Gaussian wavepacket excitation as simplest possible setup in subsection \ref{sec:gaussian_wavepacket}. For easier experimental implementation we also consider a moving Gaussian wavepacket in subsection \ref{sec:counterpropagating_wavepacket}. In the last subsection \ref{sec:boundary_wavepacket}, we implement boundary conditions and account for the influence of these boundaries in the simulation.

\subsection{Gaussian wavepacket excitation\label{sec:gaussian_wavepacket}}

\subsubsection{Initial condition}

It is known\cite{Grobe_1999_Numerical_Dirac_equation_Zitterbewegung}, that the Zitterbewegung only shows up for a simultaneous excitation of the positive eigen energy spectrum (above the mass gap of $m_0 c^2$) and the negative eigen energy spectrum (below the mass gap of $m_0 c^2$). Therefore, the simplest initial condition for a wavepacket, which exhibits Zitterbewegung dynamics is two Gaussian wavepackets in momentum space: One Gaussian wavepacket for the positive energy eigenstates and one Gaussian wavepacket for the negative energy eigenstates. Both Gaussians should be centered at momentum $k=0\,\textrm{m}^{-1}$. This description corresponds to the initial condition of the wavepacket
\begin{subequations}
\begin{align}
 \phi_k^+ &= e^{- \left(\frac{k}{\sigma_k}\right)^2}\,,\\
 \phi_k^- &= e^{- \left(\frac{k}{\sigma_k}\right)^2}\,,
\end{align}\label{eq:gaussian_wavepacket}%
\end{subequations}
at time $t=0\,\textrm{ns}$, where the positive- and negative energy eigenstates are excited equally. The width of the wavepacket should be of the order of the Compton wavelength $\hbar/(m_0 c)$, which is the typical scale of the Zitterbewegung. By applying the metamaterial scaling \eqref{eq:scaling_replacements}, this turns into $\hbar/(m' c_D)$. Thus we choose the width in frequency space to  be $\sigma_k=m' c_D \sqrt{2}/\hbar$. Correspondingly, the wavefunction \eqref{eq:wavefunction_expansion_vacuum} takes the form
\begin{equation}
 \varphi(x) = \frac{1}{\sqrt{N}} \int_{-\infty}^{\infty} dk \left(u_k^+ e^{i k x} + u_k^- e^{i k x}\right) e^{- \left(\frac{k}{\sigma_k}\right)^2}\,, \label{eq:initial_wavepacket_nonmoving}
\end{equation}
with norm
\begin{equation}
 N = \sqrt{2 \pi} \sigma_k\,. \label{eq:wavepacket_norm}
\end{equation}

\subsubsection{Simulation}

The simulation is carried with a resolution of 401 grid points in momentum space. Therefore the simulation area has an extension of $3.2\,\textrm{m}$ from the $1^\textrm{st}$ till the $401^\textrm{th}$ index, according to the numerical considerations section.
%

\begin{figure}
\begin{center}
\includegraphics[width=0.48\textwidth]{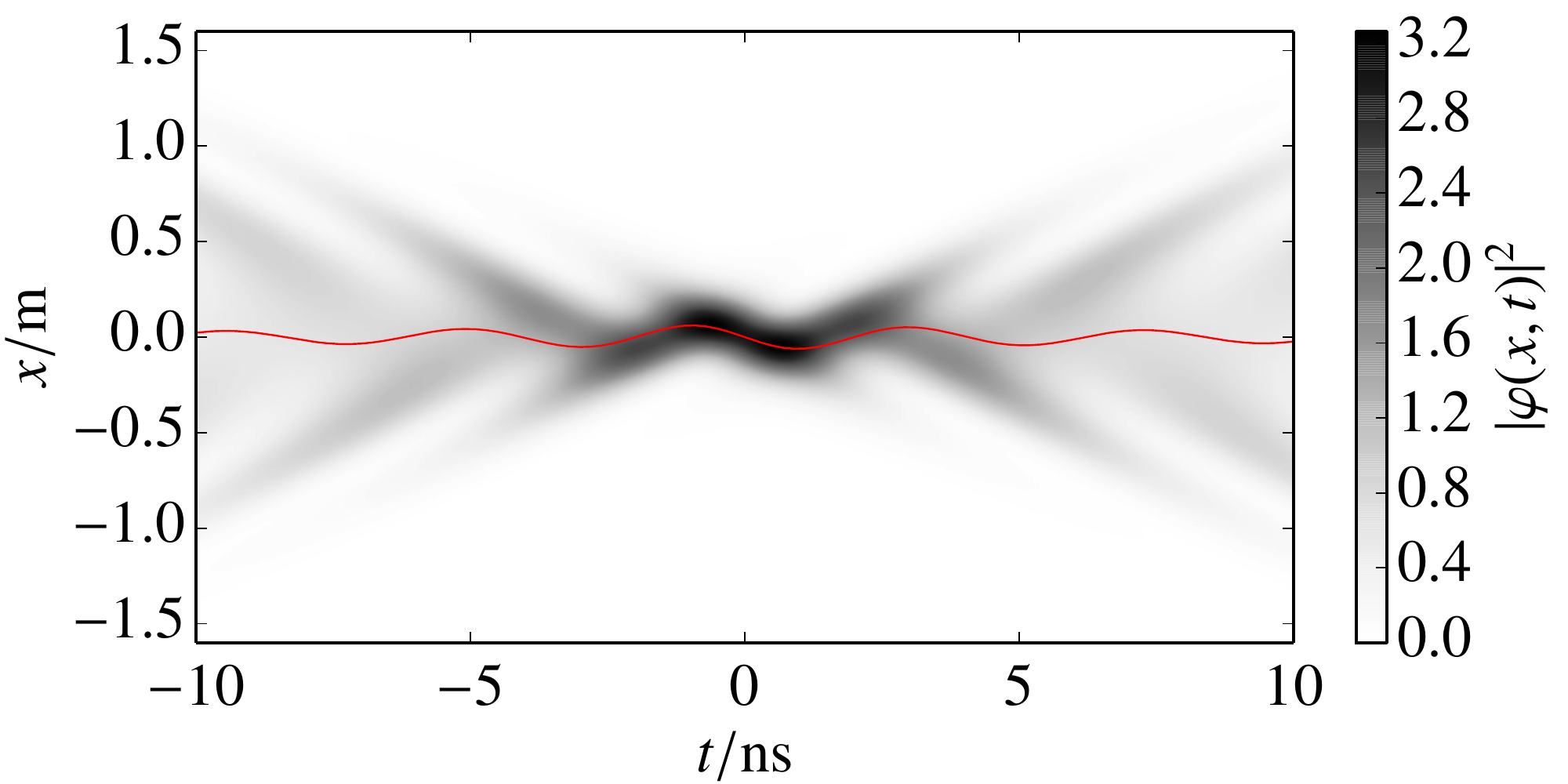}
\end{center}
\caption{
Time evolution of a Gaussian wavepacket. The figure shows the probability density of the metamaterial simulation of the effective Dirac theory according to the dynamical evolution equation \eqref{eq:wavefunction_expansion} with the initial condition \eqref{eq:gaussian_wavepacket}, which is exhibiting a Zitterbewegung. Note, that the initial condition is specified at time $t=0\,\textrm{ns}$, which means that we show the computed backward time-evolution ($t<0\,\textrm{ns}$), being in one line with the forward time-evolution ($0\,\textrm{ns}<t$). The red line displays the position expectation \eqref{eq:position_expectation_simple} and is also plotted as solid line in figure \ref{fig:simple_zitterbewegung_probability_amplitude}.
}
\label{fig:simple_zitterbewegung_probability_density}
\end{figure}

We have plotted the probability density of the metamaterial simulation with initial condition \eqref{eq:gaussian_wavepacket} in figure \ref{fig:simple_zitterbewegung_probability_density}, in which we also compute the position expectation value
\begin{equation}
 \Braket{\varphi|x(t)|\varphi} = \int_{-\infty}^\infty dx \, x \, |\varphi(x,t)|^2 \label{eq:position_expectation_simple}
\end{equation}
of the simulated probability density $|\varphi(x,t)|^2$ (red line). Equation \eqref{eq:position_expectation_simple} thereby shows the general formula for the position expectation, in which the limits of the infinite integration interval $[-\infty,\infty]$ are constrained down to the interval $[-1.6\,\textrm{m},1.6\,\textrm{m}]$ in the case of the numerical simulation. For further investigation, we plot the red line of figure \ref{fig:simple_zitterbewegung_probability_density} again in figure \ref{fig:simple_zitterbewegung_probability_amplitude} as solid black line. We compare this function with the position expectation value which is computed by using the simulated probability density of the exact Dirac equation (dotted line). The term `exact Dirac equation' means, that the spinors $u_k^+$ and $u_k^-$ are replaced by the spinors $\tilde u_k^+$ and $\tilde u_k^-$ of equation \eqref{eq:spinors_vacuum} and the frequencies $\omega_+$ and $\omega_-$ are replaced by $\pm \tilde{\mathcal{E}}(k)/\hbar$ of equation \eqref{eq:relativistic_energy-momentum-relation} in the time evolution equation \eqref{eq:wavefunction_expansion} of the Dirac equation. Another graph (plus-marked line) in figure \ref{fig:simple_zitterbewegung_probability_amplitude} is the position expectation value, which is derived analytically in appendix \ref{sec:zitterbewegung_of_real_electron}. The final equation \eqref{eq:zitterbewegung_expectation} can be adapted to our problem by inserting the initial condition \eqref{eq:gaussian_wavepacket} and applying the scaling rule \eqref{eq:scaling_replacements}, which yields the semi-analytic equation
\begin{equation}
 \Braket{\varphi|x(t)|\varphi} = \frac{1}{N} \int_{-\infty}^{\infty} dk \frac{m' \hbar c_D^3 }{{\mathcal{E}(k)}^2} e^{-2 \left(\frac{k}{\sigma_k}\right)^2} \sin\left(- \frac{2 \,\mathcal{E}(k) t}{\hbar}\right)\,.\label{eq:gaussian_zitterbewegung_expectation}
\end{equation}
Note, that we have inserted a minus sign in the argument of the sine function by hand, according to the negative time-evolution of the metamaterial simulation, which we have discussed at the end of subsection \ref{sec:scaled_dirac_equation}.

\subsubsection{Discussion}

\begin{figure}
\begin{center}
\includegraphics[width=0.48\textwidth]{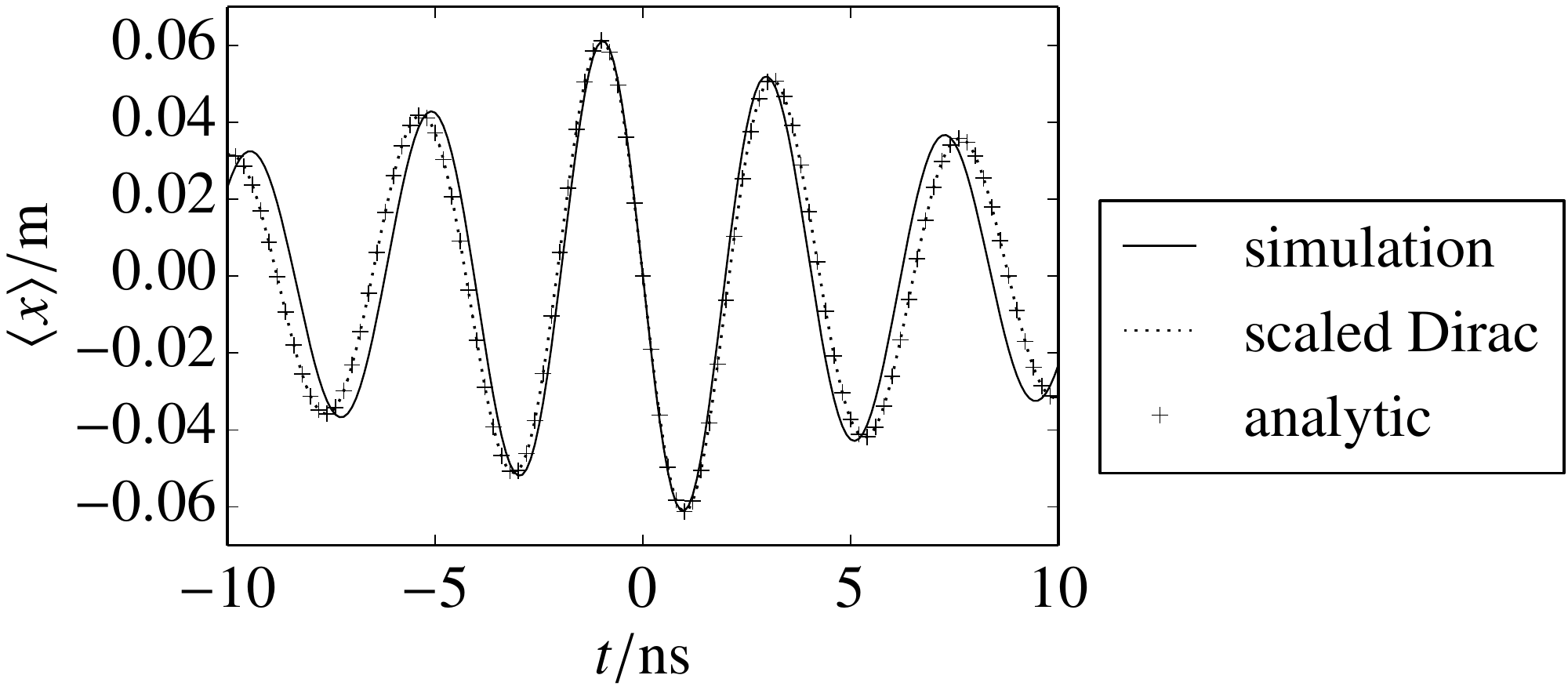}
\end{center}
\caption{
Zitterbewegung of a metamaterial simulation. We show the position expectation value \eqref{eq:position_expectation_simple} of the metamaterial simulation of the effective Dirac equation with initial condition \eqref{eq:gaussian_wavepacket}, which is illustrated in figure \ref{fig:simple_zitterbewegung_probability_density}. The solid line is an identical plot as the red line \ref{fig:simple_zitterbewegung_probability_density}. Furthermore, the dotted line results from a simulation which is based on exact Dirac theory, as described in the main text. The plus marked line is a plot of the Zitterbewegung, given by the semi-analytic solution \eqref{eq:gaussian_zitterbewegung_expectation}.
}
\label{fig:simple_zitterbewegung_probability_amplitude}
\end{figure}

The line `\emph{scaled Dirac}' and the line `\emph{analytic}' in figure \ref{fig:simple_zitterbewegung_probability_amplitude} are both based on exact Dirac theory. For this reason, they are on top of each other and therewith identical.

Furthermore, the comparison of the metamaterial simulation of the effective Dirac equation with the exact Dirac simulation in figure \ref{fig:simple_zitterbewegung_probability_amplitude} yields good agreement.
The match of the `\emph{simulation}' line with the `\emph{scaled Dirac}' line and the `\emph{analytic}' line is one of our main results. It implies, that our effective dynamic Dirac theory appears in metamaterial simulations such that the well-known Zitterbewegung can emerge. We conclude that metamaterials are capable of emulating the time-dependent Dirac equation in the case of well-suited parameters as for this setup.

\subsection{Counterpropagating excitation\label{sec:counterpropagating_wavepacket}}

\subsubsection{Initial condition}

The Gaussian wavepacket excitation in the above subsection is entering and exiting the simulation region very slowly, which means that the required electro-magnetic input pulse at the metamaterial interfaces will have an infinitely long head and tail. This is no useful property for an experimental realization and therefore, we demand a wavepacket which performs a Zitterbewegung but evolves more suitable in a prospective experiment. This property can be achieved by shifting the Gaussian wave packet \eqref{eq:gaussian_wavepacket} to the right in momentum space by the value $k_0 = 20\,\textrm{m}^{-1}$, implying the new initial condition
\begin{subequations}
\begin{align}
 \phi_k^+ &= e^{- \left(\frac{k - k_0}{\sigma_k}\right)^2}\,, \\
 \phi_k^- &= e^{- \left(\frac{k - k_0}{\sigma_k}\right)^2}\,,
\end{align}\label{eq:initial_condition_colliding}%
\end{subequations}
at time $t=0\,\textrm{ns}$. Here, the width of the wavepacket in frequency space is chosen to be  $\sigma_k=m' c_D/\hbar$. Consequently, the wavefunction \eqref{eq:initial_wavepacket_nonmoving} changes into
\begin{equation}
 \varphi(x) = \frac{1}{\sqrt{N}} \sum_k \left(u_k^+ e^{i k x} + u_k^- e^{i k x} \right) e^{- \left(\frac{k - k_0}{\sigma_k}\right)^2}\,, \label{eq:initial_wavepacket_colliding}
\end{equation}
where the norm \eqref{eq:wavepacket_norm} remains unchanged, except of course, the change of $\sigma_k$.

\subsubsection{Simulation}

For the new setup, we increase the number of spacial grid points to 625.
As a consequence, the simulation region is now extended over the interval $[-2.496\,\textrm{m},2.496\,\textrm{m}]$.

\begin{figure*}
\begin{center}
\includegraphics[width=1.0\textwidth]{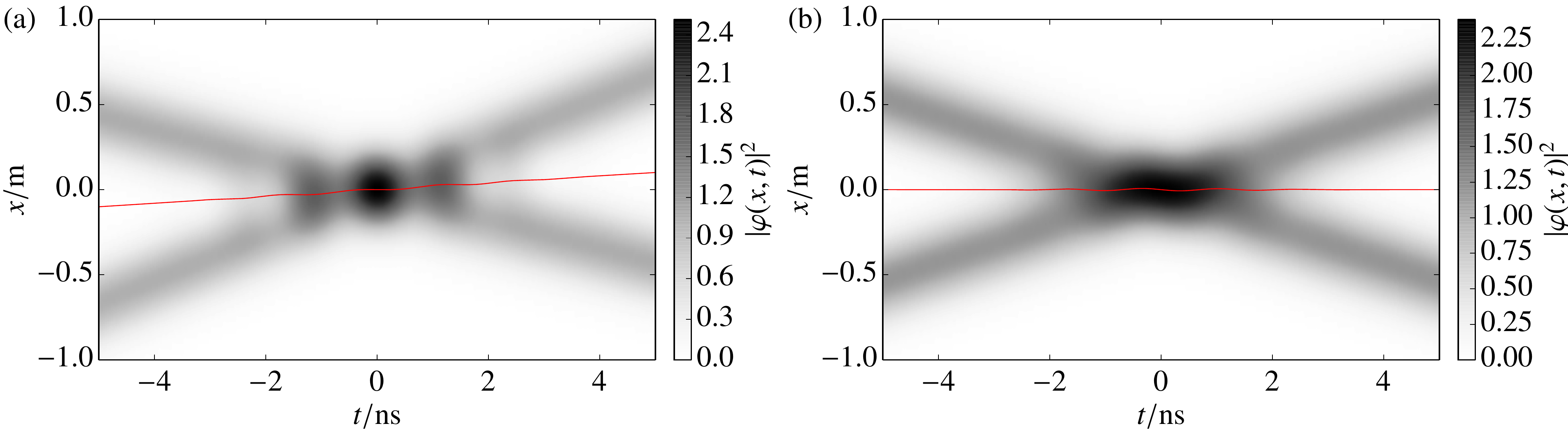}
\end{center}
\caption{
Time evolution of a colliding wavepacket. (a) The metamaterial time-evolution according to equation \eqref{eq:wavefunction_expansion} with initial condition \eqref{eq:initial_condition_colliding} The red line marks the computed position expectation $\Braket{x(t)}$ of \eqref{eq:position_expectation_simple}, which is also shown as solid line in figure \ref{fig:collision_zitterbewegung_probability_amplitude} (a).\quad(b) The time-evolution of the same initial condition, but simulated with exact Dirac theory. The corresponding position expectation (red line) is also plotted as black dotted line in figure \ref{fig:collision_zitterbewegung_probability_amplitude} (b). Like in figure \ref{fig:simple_zitterbewegung_probability_amplitude}, the initial condition for the simulation is specified at $t=0\,\mathrm{ns}$, from which the backward and forward time-evolution is computed.
}
\label{fig:collision_zitterbewegung_probability_density}
\end{figure*}

Like in the above subsection \ref{sec:gaussian_wavepacket} we have plotted the time-evolution of the probability density of the metamaterial simulation for the initial condition \eqref{eq:initial_condition_colliding} in figure \ref{fig:collision_zitterbewegung_probability_density}. One can see, that the positive and negative energy eigenstates are counterpropagating from $x=-\infty$ and $x=\infty$ at time $t=-\infty$, collide at $x=0$ at time $t=0$ and move apart from each other to $x=-\infty$ and $x=\infty$ at time $t=+\infty$. An oscillatory pattern is appearing at the colliding point (see also figure \ref{fig:collision_zitterbewegung_probability_amplitude}), which we attribute as Zitterbewegung dynamics. The reader might expect, that positive and negative states move in the same direction, because we have shifted the excitation by momentum $k_0$ to the right in momentum space. But since the upper and lower band of the dispersion relation have the opposite slope at $k_0$ the shift in momentum space results in two counterpropagating wave packets.

The metamaterial simulation of the effective Dirac equation and the simulation of the exact Dirac equation differ significantly from each other. Therefore, we plot them in two different plots (a) and (b) in figure \ref{fig:collision_zitterbewegung_probability_density}, respectively.
In the case of the effective Dirac theory of the metamaterial, there is an asymmetry which we explain with the asymmetry of the dispersion relation (see figure \ref{fig:dispersion_relation} (a)). The positive energy eigenstates are moving with group velocity $(\partial \omega_- / \partial k) |_{k_0}$ at $k_0$ of the lower band, while the negative energy eigenstates are moving with group velocity $\partial (\omega_+ / \partial k) |_{k_0}$ at $k_0$ of the upper band. Since the upper and lower band are differently shaped, the positive and negative energy-eigenstates are propagating at different speed through the medium. On the other hand, in the case of the exact Dirac equation in figure \ref{fig:collision_zitterbewegung_probability_density} (b), the dynamics appears symmetric, which we explain with the symmetric upper and lower band of the relativistic energy-momentum relation in figure \ref{fig:dispersion_relation} (b). As a result of the differences of the dispersion relations, there is an additional effective motion superimposed to the Zitterbewegung  (see red line in figure \ref{fig:collision_zitterbewegung_probability_density} (a) and solid black line in figure \ref{fig:collision_zitterbewegung_probability_amplitude} (a)), as compared to the wavepacket in the exact Dirac theory (see red line in figure \ref{fig:collision_zitterbewegung_probability_density} (b) and dotted black line in figure \ref{fig:collision_zitterbewegung_probability_amplitude} (b)).

Similarly as in the above subsection \ref{sec:gaussian_wavepacket}, we want to verify the exact Dirac theory by comparison with the semi-analytic solution in appendix \ref{sec:zitterbewegung_of_real_electron}. The new initial condition \eqref{eq:initial_condition_colliding}, substituted in equation \eqref{eq:zitterbewegung_expectation} yields
\begin{equation}
\Braket{\varphi|x(t)|\varphi} = \frac{1}{N} \int_{-\infty}^{\infty} dk \frac{m' \hbar c_D^3}{{\mathcal{E}}(k)^2} e^{- 2 \left(\frac{k - k_0}{\sigma_k}\right)^2} \sin\left(- \frac{2 \,\mathcal{E}(k) t}{\hbar}\right)\,.\label{eq:colliding_analytic_Zitterbewegung}
\end{equation}
Note, that we inserted in a minus sign by hand in the argument of the sine function, as it has been done for equation \eqref{eq:gaussian_zitterbewegung_expectation} already. We plot the value of the solution \eqref{eq:colliding_analytic_Zitterbewegung} as plus-marked line in figure \ref{fig:collision_zitterbewegung_probability_amplitude} (b).

\subsubsection{Discussion}

\begin{figure}
\begin{center}
\includegraphics[width=0.485\textwidth]{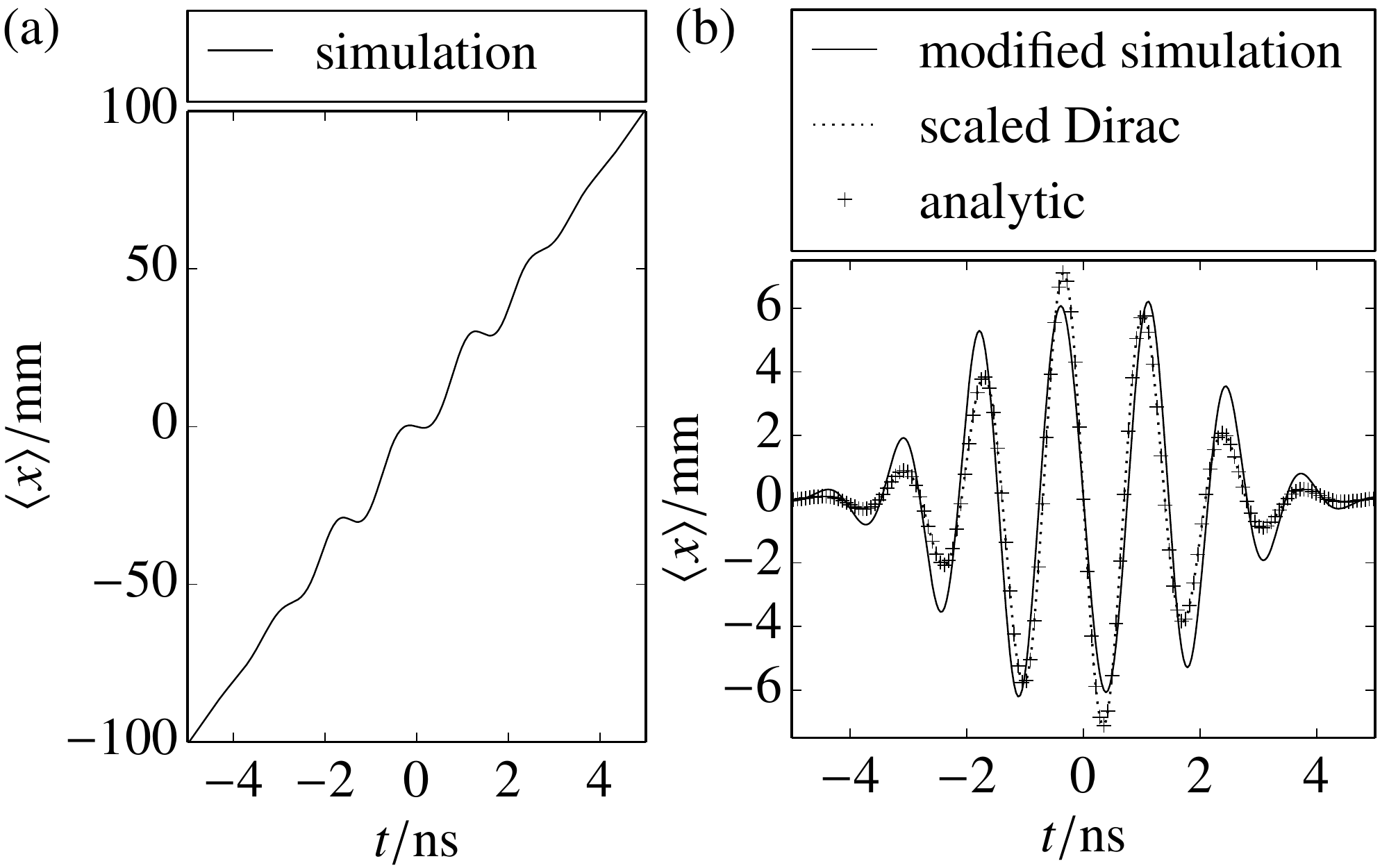}
\end{center}
\caption{
Zitterbewegung of colliding wavepackets. The solid black line in (a) and the dotted black line in (b), are the red lines in figure \ref{fig:collision_zitterbewegung_probability_density} (a) and (b) respectively. The black cross-marked line shows the data from the semi-analytic formula of the position expectation value \eqref{eq:colliding_analytic_Zitterbewegung}. For a comparison with the metamaterial simulated Dirac equation, we subtract the black solid line in (a) by $t/(49.7\,\textrm{ns})$ and plot it together with the two other lines in (b). Once the steady motion is subtracted, the metamaterial simulation and the exact Dirac theory simulation agree with each other.
}
\label{fig:collision_zitterbewegung_probability_amplitude}
\end{figure}

The lines `\emph{exact Dirac}' of the exact Dirac simulation and the `\emph{analytic}' solution in figure \ref{fig:collision_zitterbewegung_probability_amplitude} (b) are coinciding. This is to expect, because the analytic solution is based on exact Dirac theory and is a validation of both descriptions.

For a comparison of the metamaterial simulation with the exact Dirac theory, we subtract the value $t/(49.7\,\textrm{ns})$ from the position expectation value in figure \ref{fig:collision_zitterbewegung_probability_amplitude} (a), and plot it as the solid black line `\emph{modified simulation}' in figure \ref{fig:collision_zitterbewegung_probability_amplitude} (b). We obtain very good agreement of both descriptions and take this match as further confirmation for the reliability of the simulation of the effective Dirac dynamics in the metamaterial.

\subsection{Excitation input at the boundaries\label{sec:boundary_wavepacket}}

\subsubsection{Initial condition}

In the case of a metamaterial with physical boundaries, the initial wavefunction \eqref{eq:initial_wavepacket_colliding} has to be replaced by an electric and magnetic field, which is propagating from the regions (1) and (3) into region (2) of the metamaterial. This can be done by specifying the incoming electric field of the regions (1) and (3) at the boundaries by
\begin{subequations}
\begin{align}
 E_z^{(1)}(x_a,t) &= \hat E \, e^{-i \omega_a t} e^{-\left(\frac{\sigma_\omega t}{2}\right)^2} \,\label{eq:initial_wavepacket_boundaries_left} \,,\\
 E_z^{(3)}(x_b,t) &= \hat E \, e^{-i \omega_b t} e^{-\left(\frac{\sigma_\omega t}{2}\right)^2} \,.\label{eq:initial_wavepacket_boundaries_right}
\end{align}\label{eq:initial_wavepacket_boundaries}%
\end{subequations}
with an arbitrary electric field amplitude, which is chosen to be $\hat E = 1\,\textrm{V/m}$, the frequency width $\sigma_\omega = 0.52\,\textrm{GHz}$ and the two frequencies $\omega_a = 13.81\,\textrm{GHz}$ and $\omega_b = 8.95\,\textrm{GHz}$. Here, the frequency $\omega_a$ is above the band gap and $\omega_b$ is below the band gap, corresponding to the positive and negative excitation \eqref{eq:initial_condition_colliding}.

The expansion coefficients of the right propagating input pulse \eqref{eq:initial_wavepacket_boundaries_left} in region (1) are determined by the inverse Fourier transform
\begin{equation}
 E_{z,k}^{\prime(1)}(\omega) = \frac{1}{2 \pi} \int_{- \infty}^\infty dt \, E_{z}^{(1)}(x_a,t) e^{i(k x_a + \omega t)}\label{eq:time_fourier_transform}\,,
\end{equation}
in time, with corresponding Fourier transform
\begin{equation}
 E_{z}^{(1)}(x_a,t) = \int_{- \infty}^\infty d\omega \, E_{z,k}^{\prime(1)}(\omega) e^{i(k x_a - \omega t)}\label{eq:frequency_fourier_transform}
\end{equation}
in frequency space. However, the equidistant grid of the simulated time evolution \eqref{eq:wavefunction_expansion} in momentum space implies, that the sum of the Fourier transform has to be expressed in terms of an integral over the variable $k$ instead of $\omega$. Such a momentum space integral of \eqref{eq:frequency_fourier_transform} in the fashion of \eqref{eq:field_expansion} would read as
\begin{equation}
 E_z^{(1)}(x,t)=\int_0^\infty dk \Bigg[ E_{z,k}^{(1)}(\omega) e^{i(- k x - \omega t)} + E_{z,k}^{(1)}(-\omega) e^{i(- k x + \omega t)} \Bigg]\,, \label{eq:momentum_fourier_transform}
\end{equation}
where the sum with momentum space measure $\Delta k$ is replaced by an integral over $dk$. We consider a right propagating input pulse, therefore only right propagating modes with $0 \le k$ are accounted for. Since we want to adapt the same density of states in region (1) as in region (2) in the numerical implementation, we relate frequency space to momentum space by the dispersion relation \eqref{eq:dispersion-relation} in the Fourier transform. Accordingly, the function $\omega(k)$ implies an inner derivative in the integral
\begin{equation}
 E_z^{(1)}(x,t)=\int_0^\infty d k \left| \frac{\partial \omega}{\partial k} \right| \Bigg[ E_{z,k}^{\prime(1)}(\omega) e^{i(- k x - \omega t)}
 + E_{z,k}^{\prime(1)}(-\omega) e^{i(- k x + \omega t)} \Bigg]\,, \label{eq:momentum_fourier_transform_iner_derivative}
\end{equation}
and by comparison with \eqref{eq:momentum_fourier_transform}, one obtains the relation
\begin{equation}
 E_{z,k}^{(1)}(\omega) = \left| \frac{\partial \omega}{\partial k} \right| E_{z,k}^{\prime(1)}(\omega) \  \Leftrightarrow \   E_{z,k}^{\prime(1)}(\omega) = \left| \frac{\partial k}{\partial \omega} \right| E_{z,k}^{(1)}(\omega)\,.
\end{equation}
By these relations, the expansion coefficients of the Fourier transform in time and frequency space, which is required for obtaining expansion coefficients of the input pulse \eqref{eq:initial_wavepacket_boundaries_left} can be related to the Fourier transform between momentum and position space \eqref{eq:field_expansion}, which is required for the metamaterial simulation and accordingly the Dirac time-evolution \eqref{eq:wavefunction_expansion}.

The same considerations also hold in an analogous way for the left propagating input pulse \eqref{eq:initial_wavepacket_boundaries_right} with expansion coefficients $E_{z,-k}^{(3)}(\omega)$ in region (3), as well as for the right propagating reflected pulse with expansion coefficients $E_{z,k}^{(3)}(\omega)$ in region (3) and the left propagating reflected pulse $E_{z,-k}^{(1)}(\omega)$ in region (1). Note, that in the expressions \eqref{eq:momentum_fourier_transform} and \eqref{eq:momentum_fourier_transform_iner_derivative} the integral over $k$ is replaced by a sum over an equidistant momentum grid with momentum space spacing length $\Delta k$ in the numerical implementation.

Within the framework of the herein considered integral measures, the two reflected `output' pulses and the electric field inside of the metamaterial are determined for each momentum, ie. each frequency according to the system of equations of the boundary conditions \eqref{eq:explicit_boundary_conditions_left_E}, \eqref{eq:explicit_boundary_conditions_right_E} and \eqref{eq:explicit_boundary_conditions_II} of section \ref{sec:boundary_conditions}. The fully determined electric, and therewith magnetic field in region (2) in form of the expansion coefficients $E_{z,k}^{(2)}$ and $H_{y,k}^{(2)}$ can be used to compute the wavefunction's expansion coefficients $\phi_k$ by making use of the relations \eqref{eq:expansion_coefficient_equality}. Once the wavefunction is determined by this procedure, we normalize it by 
\begin{equation}
 N=\int_{-\infty}^\infty \left[ |\phi_k^+(\omega_-)|^2 + |\phi_k^-(\omega_+)|^2 \right] \Delta k\,,\label{eq:momentum_space_normalization}
\end{equation}
which according to Parseval's theorem, is equivalent to normalization in position space \eqref{eq:position_space_normalization}. In contrast to the initial condition at time $t=0$ along the whole $x$-axis in the above two subsections, this most realistic simulation specifies the `initial condition' at the boundary positions $x_a=-1.0\,\textrm{m}$ for \eqref{eq:initial_wavepacket_boundaries_left} and $x_b=1.0\,\textrm{m}$ for \eqref{eq:initial_wavepacket_boundaries_right} along the time-axis. However, through the relations (\ref{eq:left-moving_field_expansion},\ref{eq:right-moving_field_expansion},\ref{eq:signal_displacement}), the alignment of the `initial condition' can be recast from the time-axis to the $x$-axis regions (1) and (3) at a certain time $t \ll 0\,\textrm{ns}$.

\subsubsection{Simulation}

For the setup of this simulation we use $1001$ grid points in position and momentum space, which implies that the simulation is extended on the interval $[-4\,\textrm{m},4 \textrm{m}]$. For this setup, the situation of figure \ref{fig:periodic_boundary_conditions} applies, in which the boundary conditions of the metamaterial are implemented inside of the simulation interval, in which the physical metamaterial (corresponding to region (2)) is extended from $x_a$ to $x_b$ on the interval $[-1\,\textrm{m},1\,\textrm{m}]$. The intervals $[-4\,\textrm{m},-1\,\textrm{m}]$ and $[1\,\textrm{m},4\,\textrm{m}]$ correspond to  unphysical regions, which are depicted as dump space in figure \ref{fig:periodic_boundary_conditions}. Therefore, even though the time-evolution has periodic boundary conditions, the wavepacket does not immediately reenter the simulation region (2) from the other side, after it went out before. 

\subsubsection{Discussion}

\begin{figure}
\begin{center}
\includegraphics[width=0.48\textwidth]{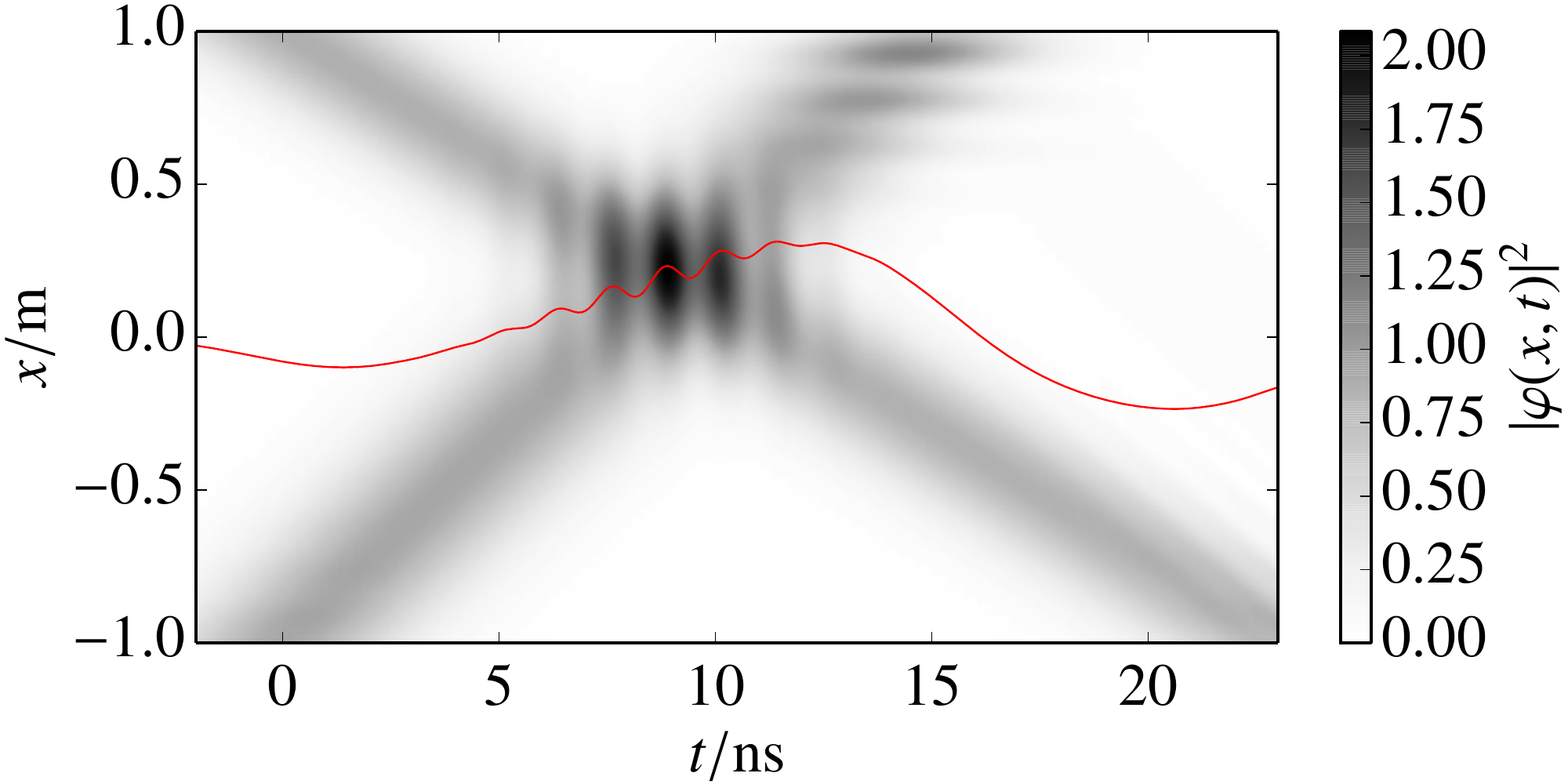}
\end{center}
\caption{
Time-evolution with boundary conditions. This figure shows a similar time-evolution as in figure \ref{fig:collision_zitterbewegung_probability_density} (a), but this time, the wavefunction is given by the boundary electric field \ref{eq:initial_wavepacket_boundaries} at the two boundaries $x_a=-1\,\textrm{m}$ and $x_b=1\,\textrm{m}$ of the metamaterial. The wavepacket's expansion coefficients and the time-evolution in region (2) are determined according to the procedure in subsection \ref{sec:time-evolution_dirac-equation} and section \ref{sec:boundary_conditions}. The position expectation value \eqref{eq:position_expectation_simple} of this wavepacket is computed on the interval $[x_a,x_b]$ of region (2) and plotted as overlayed red line and in a magnified version in figure \ref{fig:boundary_zitterbewegung_probability_amplitude}.
}
\label{fig:boundary_zitterbewegung_probability_density}
\end{figure}

\begin{figure}
\begin{center}
\includegraphics[width=0.48\textwidth]{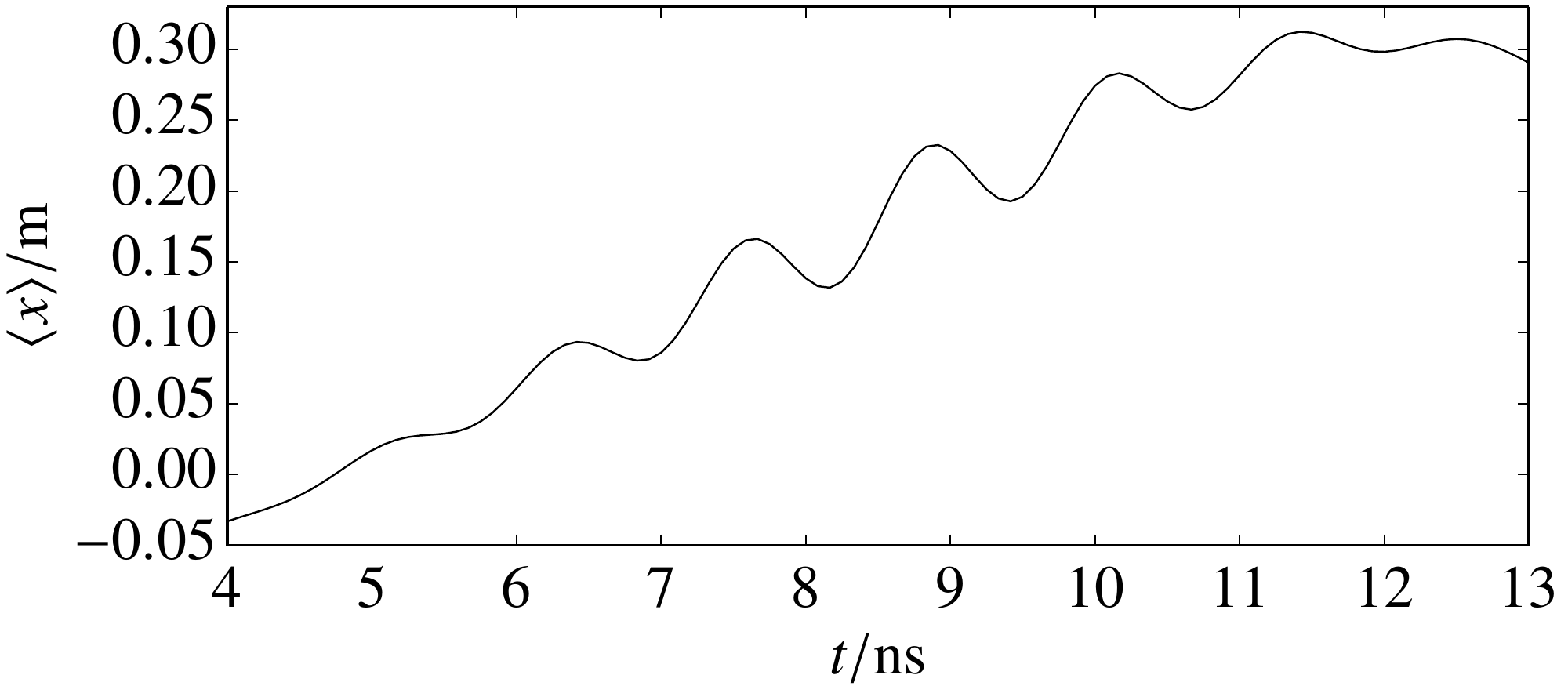}
\end{center}
\caption{
Zitterbewegung with boundary conditions. The plotted graph is the position expectation value, which is computed as red line in the boundary specified simulation in figure \ref{fig:boundary_zitterbewegung_probability_density}. Here, we show a magnification of the region, in which the excitation of the positive energy-eigenstates coincides with the excitation of the negative energy-eigenstates in space. Like in figure \ref{fig:collision_zitterbewegung_probability_amplitude}, an oscillatory pattern shows up, which we attribute as Zitterbewegung.
}
\label{fig:boundary_zitterbewegung_probability_amplitude}
\end{figure}

The simulation of a Zitterbewegung with real metamaterial boundaries is shown in figure \ref{fig:boundary_zitterbewegung_probability_density} and the position expectation value of this simulation is overlayed as red line and plotted a second time in figure \ref{fig:boundary_zitterbewegung_probability_amplitude}. A similar simulation for the exact Dirac equation with the same initial condition differs much from the effective Dirac dynamics of the metamaterial (see also figure \ref{fig:collision_zitterbewegung_probability_density}) and avoids the feasibility of a direct comparison. Therefore, there is no comparison with the corresponding dynamics of the exact Dirac equation in this subsection.


In figure \ref{fig:boundary_zitterbewegung_probability_density} one can see, that the electro-magnetic excitation in the setup is propagating from the physical boundaries towards the central region of the simulation area. In the overlap region of both excitations an oscillatory pattern emerges (see figure \ref{fig:boundary_zitterbewegung_probability_amplitude}) which we attribute as Zitterbewegung of the effective Dirac equation in the metamaterial. This demonstrates, that in the case of an ideal metamaterial, given by the equations \eqref{eq:maxwell_equations} and \eqref{eq:permittivity_and_permeability} a Zitterbewegung of a simulated Dirac equation may occur for a real experiment.

\section{Conclusion and Outlook}

The article discusses the feasibility on an exactly mappable, time-dependent Dirac simulation, which is actually emulated by electro-dynamics of Maxwell equations with designed electro-magnetic properties of a metamaterial structure. We present a time-evolution equation in a Dirac-like fashion \eqref{eq:wavefunction_expansion} and give arguments of why it evolves very similar to exact Dirac theory. Furthermore, we consider scaling relations in subsection \ref{sec:scaled_dirac_equation}, with which the slower dynamics of the metamaterial can be mapped to the faster dynamics of elementary particles in Dirac theory. Based on the scaling rules and the exact Dirac equation, we have derived an explicit, semi-analytic expression for the position expectation of a general wavefunction in appendix \ref{sec:zitterbewegung_of_real_electron}.

We demonstrate the dynamic description with three simulations: In subsection \ref{sec:gaussian_wavepacket} and \ref{sec:counterpropagating_wavepacket}, we compare our metamaterial simulations with exact Dirac theory and with the analytic solution, which is derived within this exact Dirac theory. We get very good agreement between the emulated Dirac dynamics and the exact Dirac dynamics, which we interpret as validation of our description. It is also a proof for the emergence of  Zitterbewegung, which is a characteristic property of the free Dirac equation. Subsequently, we present a more realistic scenario, in which boundary conditions are included, such that the simulation can be realized in experiment in subsection \ref{sec:boundary_wavepacket}. In this most realistic simulation we also observe an oscillatory pattern, which matches the Zitterbewegung.

We conclude that a Dirac wavefunction can be emulated by the electro-magnetic field in a metamaterial, such that the Zitterbewegung occurs. Next steps are an experimental implementation of this theoretical study of the effect and also an investigation of the quantization of the system.

\begin{acknowledgments}
S. A. would like to thank Hong Chen for the support nice hospitality of the collaborative visits at the Tongji University. S. A. also gratefully thanks for the helpful discussions about Zitterbewegung and periodic boundery conditions with Rainer Grobe from the Illinois State University.
\end{acknowledgments}

\appendix

\section{Zitterbewegung of a real electron in Dirac theory\label{sec:zitterbewegung_of_real_electron}}

For verifying exact Dirac theory simulations, we consider the time-dependent position operator $x(t)$ of the free Dirac equation in the Heisenberg picture in this section. This is already discussed in the literature. We simplify the already known derivation\cite{Schroedinger_1930_Zitterbewegung,Milonni_1994_Quantum__Vacuum,Grobe_1999_Numerical_Dirac_equation_Zitterbewegung} of the position operator for the one-dimensional case. Applying a general wavefunction to this position operator, yields a semi-analytic formula for the computation of the wavepacket's position expectation value.

We start out with the velocity operator $c \sigma_x$ in the Heisenberg picture
\begin{equation}
 \partial_t x = \frac{i}{\hbar} \left[ H,x \right] = c \sigma_x\,,
\end{equation}
where $H$ is the Hamiltionan of the exact Dirac theory \eqref{eq:vacuum_dirac_hamiltonian}.
A second application of the Heisenberg equation yields the acceleration operator
\begin{equation}
 \partial_t c \sigma_x = \frac{i}{\hbar} \left[ H, c \sigma_x \right] = - \frac{2 c \sigma_y m_0 c^2}{\hbar} = \frac{2 i c}{\hbar} \left( c p_x - \sigma_x H \right)\,.
\end{equation}
If one accounts for the time-independent constants $H$ and $p_x$, one finds the formal solution
\begin{equation}
 c \sigma_x(t) = \left( c \sigma_x(0) - c^2 p_x H^{-1} \right) e^{-i 2 H t/\hbar} + c^2 p_x H^{-1}\,.
\end{equation}
Integration of the velocity operator with respect to time gives the solution of the time-dependent position operator in the Heisenberg picture
\begin{equation}
 x(t) = x_0 + c^2 p_x H^{-1} t
 - \frac{i}{2} \hbar c \left( \sigma_x H^{-1} - c p_x H^{-2} \right) \left( 1 - e^{- i 2 H t/\hbar} \right)\,.\label{eq:time-dependent_position_space_operator}
\end{equation}
In the following, we deduce the expectation value of this operator for a general wavefunction
\begin{equation}
 \varphi(x,t) = \frac{1}{\sqrt{N}} \int_{-\infty}^{\infty} \left(\phi_k^+ \, \tilde u_k^+ \, e^{i k x} + \phi_k^- \, \tilde u_k^- \, e^{i k x} \right) dk\,.\label{eq:wavefunction_expansion_vacuum}
\end{equation}
Note, that in contrast to \eqref{eq:wavefunction_expansion} the spinors no longer have a frequency dependent mass in the Dirac theory of an elementary particle. Accordingly, due to the equations of motion \eqref{eq:quantum_mechanical_time_evolution} and \eqref{eq:vacuum_dirac_hamiltonian}  the energy-momentum relation
\begin{equation}
 \tilde{\mathcal{E}}(k) = \sqrt{c^2 \hbar^2 k^2 + m_0^2 c^4}\label{eq:relativistic_energy-momentum-relation}
\end{equation}
and the spinors
\begin{subequations}
\begin{align}
 \tilde u_{k}^+ &=
 \frac{1}{\sqrt{2 |\tilde{\mathcal{E}}(k)|(|\tilde{\mathcal{E}}(k)| + m_0 c^2)}}
 \begin{pmatrix}
  |\tilde{\mathcal{E}}(k)| + m_0 c^2 \\ c \hbar k
 \end{pmatrix}\,,\label{eq:positive_spinor_vacuum}\\
 \tilde u_{k}^- &=
 \frac{1}{\sqrt{2 |\tilde{\mathcal{E}}(k)|(|\tilde{\mathcal{E}}(k)| + m_0 c^2)}}
 \begin{pmatrix}
  - c \hbar k \\ |\tilde{\mathcal{E}}(k)| + m_0 c^2
 \end{pmatrix}\label{eq:negative_spinor_vacuum}
\end{align}\label{eq:spinors_vacuum}%
\end{subequations}
contain the natural constants $m_0$, $c$ and $\hbar$. The computation of the expectation value of the position operator is as follows. First we define the function
\begin{equation}
 f(k) := \phi_k^+ \, \tilde u_k^+ + \phi_k^- \, \tilde u_k^-\,.
\end{equation}
With this abbreviation, the position expectation value of the position operator can be written as
\begin{equation}
 \braket{\varphi|x(t)|\varphi} = \int_{- \infty}^\infty dx \int_{- \infty}^\infty dk \, e^{-i k x} f(k)^\dagger x(t)
 \int_{- \infty}^\infty dk' e^{i k' x} f(k')\,.\label{eq:position_expectation_value_definition}
\end{equation}
The operator $x(t)$ acts at on the right hand side integral, which is the integral over $k'$. This integral is a continuous sum over the exponential functions $e^{-i k' x}$ and one can commute the operator with the right hand side integration, where it solely acts at these exponentials. Hence, the position operator \eqref{eq:time-dependent_position_space_operator} turns into a function of $k'$, which means, that all momentum operators $p_x$ of the operator $x(t)$ and $H$ in $x(t)$ turn into the number $\hbar k'$. It remains a triple integral over the function
\begin{equation}
  e^{-i (k-k') x} f(k)^\dagger x(t,k') f(k')
\end{equation}
with the three integration variables $x$, $k$ and $k'$. If one performs the integration over $x$ first, the exponential $e^{-i (k-k') x}$ turns into the delta function $2 \pi \, \delta( k- k')$. Another integration over $k'$ gives contributions only at $k=k'$, such that equation \eqref{eq:position_expectation_value_definition} turns into
\begin{equation}
 \braket{\varphi|x(t)|\varphi} = 2 \pi \int_{- \infty}^\infty dk \,f(k)^\dagger x(t,k) f(k')\,.
\end{equation}
The remaining integrand can be reshaped by making use of the eigenvalue relations
\begin{align}
 H(k)\, \tilde u_k^+ &= \phantom{-}\mathcal{E}(k)\, \tilde u_k^+ \,,\\
 H(k)\, \tilde u_k^- &= -\mathcal{E}(k)\, \tilde u_k^-\,,
\end{align}
the orthonormal property
\begin{subequations}
\begin{align}
 \tilde u_k^{+\dagger} \tilde u_k^+ &= 1\,, \\
 \tilde u_k^{+\dagger} \tilde u_k^- &= 0\,, \\
 \tilde u_k^{-\dagger} \tilde u_k^+ &= 0\,, \\
 \tilde u_k^{-\dagger} \tilde u_k^- &= 1
\end{align}
\end{subequations}
and the basic identities
\begin{subequations}
\begin{align}
 \tilde u_k^{+\dagger} \sigma_1 \tilde u_k^+ &= \frac{c \hbar k}{\tilde{\mathcal{E}}(k)}\,, \\
 \tilde u_k^{+\dagger} \sigma_1 \tilde u_k^- &= \frac{m_0 c^2}{\tilde{\mathcal{E}}(k)}\,, \\
 \tilde u_k^{-\dagger} \sigma_1 \tilde u_k^+ &= \frac{m_0 c^2}{\tilde{\mathcal{E}}(k)}\,, \\
 \tilde u_k^{-\dagger} \sigma_1 \tilde u_k^- &= - \frac{c \hbar k}{\tilde{\mathcal{E}}(k)}\,.
\end{align}
\end{subequations}
Applying these identities, canceling some terms and rearranging the remaining terms yields
\begin{widetext}
\begin{multline}
 \Braket{\varphi|x(t)|\varphi} 
 = \frac{1}{N} \int_{-\infty}^{\infty} dk \bigg\{ x_0 + \left[ |\phi^+(k)|^2 - |\phi^-(k)|^2 \right] \frac{c^2 \hbar k t}{\tilde{\mathcal{E}}(k)}\\
 + \frac{i}{2} \frac{m_0 \hbar c^3}{\tilde{\mathcal{E}}(k)^2} \left[ \phi^+(k)^* \phi^-(k) \left( 1 - e^{i 2 \tilde{\mathcal{E}}(k) t/\hbar} \right) - \phi^+(k) \phi^-(k)^* \left( 1 - e^{- i 2 \tilde{\mathcal{E}}(k) t/\hbar} \right) \right] \bigg\}\,.\label{eq:zitterbewegung_expectation_complex}
\end{multline}
With a further usage of Euler's formula this result can be further rewritten in the form
\begin{multline}
 \Braket{\varphi|x(t)|\varphi}
 = \frac{1}{N} \int_{-\infty}^{\infty} dk \bigg\{ x_0 + \left[ |\phi^+(k)|^2 - |\phi^-(k)|^2 \right] \frac{c^2 \hbar k t}{\tilde{\mathcal{E}}(k)} \\
 - \frac{m_0 \hbar c^3}{\tilde{\mathcal{E}}(k)^2} \left[ Im\left[\phi^+(k)^* \phi^-(k) \right]\left[ 1 - \cos\left(\frac{2 \tilde{\mathcal{E}}(k) t}{\hbar}\right) \right]
 - Re\left[\phi^+(k)^* \phi^-(k) \right] \sin\left(\frac{2 \tilde{\mathcal{E}}(k) t}{\hbar}\right) \right] \bigg\}\,.\label{eq:zitterbewegung_expectation}
\end{multline}
\end{widetext}

\section{The wavefunction's imaginary part in reality\label{sec:imaginary_part}}

The whole description of the simulation of the electro-magnetic waves in the metamaterial and even the basic equations \eqref{eq:maxwell_equations}, on which our theory is built on, as well as in the description of the Dirac equation makes use of complex numbers. As a result, the electric and magnetic field has an imaginary part. However, in reality nothing is imaginary and there must be an answer of what happens with this kind of `degree of freedom', which is not explicitly showing up in the real experiment.

Consider the expansion
\begin{equation}
 f(x) = \sum_k \tilde f(k) e^{i k x}\,.\label{eq:simple_expansion}
\end{equation}
The expansion will be exclusively real, if all expansion coefficients only have a symmetric real part
\begin{equation}
 Re\left(\tilde f(k)\right) = Re\left(\tilde f(-k)\right)
\end{equation}
and an anti-symmetric imaginary part
\begin{equation}
 Im\left(\tilde f(k)\right) = -Im\left(\tilde f(-k)\right)\,.
\end{equation}
On the other hand, the expansion will be exclusively imaginary, if all expansion coefficients only have an anti-symmetric real part
\begin{equation}
 Re\left(\tilde f(k)\right) = -Re\left(\tilde f(-k)\right)
\end{equation}
and a symmetric imaginary part
\begin{equation}
 Im\left(\tilde f(k)\right) = Im\left(\tilde f(-k)\right)\,.
\end{equation}
In \eqref{eq:field_expansion} the sum in the expansion runs over terms with negative frequencies and negative momenta at the same time and the discussion of exclusively real or imaginary functions in analogy to \eqref{eq:simple_expansion} gets more involved. Table \ref{tab:real_and_imaginary_wavefunction} lists the different combination of symmetries, which are possible for the expansion coefficients $E_{z,k}$ and $H_{y,k}$. Note, that the real part and the imaginary part can be chosen independently from each other. This means, that the resulting function will be only exclusively real or exclusively imaginary, if the symmetries of the real part and the symmetries of the imaginary part are both chosen either real or imaginary at the same time. According to these considerations, it is possible to choose expansion coefficients, such that the approximated function is exclusively real.
\begin{table}[!ht]
\caption{
\bf{Expansion coefficient symmetries}}
\begin{center}
\begin{tabular}{r|r|r||c}
    coefficient &      frequency &       momentum & function \\ \hline \hline
      real part &      symmetric &      symmetric & real \\
      real part &      symmetric & anti-symmetric & imaginary \\
      real part & anti-symmetric &      symmetric & imaginary \\
      real part & anti-symmetric & anti-symmetric & real \\ \hline
 imaginary part &      symmetric &      symmetric & imaginary \\
 imaginary part &      symmetric & anti-symmetric & real \\
 imaginary part & anti-symmetric &      symmetric & real \\
 imaginary part & anti-symmetric & anti-symmetric & imaginary
\end{tabular}
\end{center}
\begin{flushleft}
This table lists the different symmetries of the real part and the imaginary part of the expansion coefficients $E_{z,k}$ and $H_{y,k}$ in the sum \eqref{eq:field_expansion} with respect to frequency $\omega$ and momentum $k$, in analogous extension to the symmetry considerations of the expansion \eqref{eq:simple_expansion}. The first column `\emph{coefficient}' tells if the real part or the imaginary part of the wave function is considered, the second `\emph{frequency}' and third `\emph{momentum}' column tells, whether the real/imaginary part of the expansion coefficients are assumed to be fully symmetric or fully anti-symmetric with respect to frequency or momentum, respectively. The last line `\emph{function}' tells if the resulting function of this expansion will be exclusively real or exclusively imaginary.
\end{flushleft}
\label{tab:real_and_imaginary_wavefunction}
\end{table}

Having said that the electric and magnetic field can be exclusively real in the expansion \eqref{eq:simple_expansion}, we additionally point out, that the Maxwell equations in momentum- and frequency space \eqref{eq:maxwell_equations_frequency_space_k} are flipping the symmetry type from symmetric to anti-symmetric and vice versa with respect to frequency and momentum simultaneously, when they are coupling the electric expansion coefficients $E_{z,k}$ to the magnetic expansion coefficients $H_{y,k}$. By comparing this flipping operation with table \ref{tab:real_and_imaginary_wavefunction}, one realizes, that functions which are real, are only mapped to functions which are real again. As a consequence, it is possible to choose expansion coefficients, such that the electric and magnetic fields are real in space and time and furthermore are compatible with the Maxwell equations \eqref{eq:maxwell_equations_frequency_space_k}. So the theory, which we have developed in the subsections above is compatible with a real electric field and a real magnetic field, even though complex numbers are showing up in the equations.

However, the Dirac equation (\ref{eq:quantum_mechanical_time_evolution},\ref{eq:vacuum_dirac_hamiltonian}) explicitly implies complex wavefunctions. The time evolution \eqref{eq:wavefunction_expansion} of the effective Dirac equation can be directly translated to the electric and magnetic field by the relations \eqref{eq:wavefunction_em-field_relations} in real space and \eqref{eq:expansion_coefficient_equality} in momentum and frequency space. Therefore, the question why the electro-magnetic field does not contain an imaginary part in reality, but in contrast the Dirac wavefunction does contain an imaginary part is still to be answered!

The difference of the description of the Maxwell equations and the Dirac equation can be found in the expansions \eqref{eq:field_expansion} and \eqref{eq:wavefunction_expansion}. While the sum in \eqref{eq:wavefunction_expansion} only runs over the positive frequencies $\omega_+$ and $\omega_-$, it also runs over the negative frequencies $-\omega_+$ and $-\omega_-$ in \eqref{eq:field_expansion}. Therefore, on one hand, the simulated Dirac equation only makes use of the upper two bands of the dispersion relation \eqref{eq:dispersion-relation} and is consistent with the two bands of the positive and negative energy eigenstates of the effective  Dirac equation \eqref{eq:scaled_dirac_equation} (see also figure \ref{fig:dispersion_relation}). On the other hand, the dispersion relation \eqref{eq:dispersion-relation} has four bands, of which the two negative ones have the just the negative value of the positive ones. So the expansion of the electro-magnetic field runs additionally over these two negative energy bands. As a conclusion, the time-evolution of the Dirac equation \eqref{eq:wavefunction_expansion} omits the sum over the negative energy bands of the electro-magnetic field, without explicitly mentioning it. This omittance is introduced to be consistent with the two bands of conventional Dirac theory. However, the real- and imaginary part of the expansion coefficients $E_{z,k}(-\omega)$, $H_{y,k}(-\omega)$ of the negative energy band of the Maxwell equations can be chosen according to the symmetry scheme of table \ref{tab:real_and_imaginary_wavefunction}, such that the electric and magnetic field, which is transformed from the wavefunction by the relations \eqref{eq:expansion_coefficient_equality} becomes exclusively real. If one eliminates the imaginary part by the above discussed symmetry procedure and divides by a factor of two, one acted as if one just took the real part of the wavefunction. Therefore, the considerations of this appendix are justifying that the real part of the simulated Dirac wavefunction is proportional to the electric and magnetic field of the underlying Maxwell equations, which can be measured in experiment.



\bibliographystyle{unsrt}
\bibliography{bibliography}

\end{document}